\documentclass[aps,prc,twocolumn,unsortedaddress,showpacs]{revtex4-1}

\usepackage{graphicx}

\begin{document}


\title{Intermittency route to chaos for the nuclear billiard - a quantitative study}

\author{Daniel Felea}
\email{dfelea@spacescience.ro}
\affiliation{Institute of Space Sciences, P.O.Box MG 23, RO 77125, Bucharest-M\u{a}gurele, Romania}

\author{Cristian Constantin Bordeianu}
\author{Ion Valeriu Grossu}
\author{C\u{a}lin Be\c{s}liu}
\author{Alexandru Jipa}
\affiliation{Faculty of Physics, University of Bucharest, P.O.Box MG 11, RO 77125, Bucharest-M\u{a}gurele, Romania}

\author{Aurelian-Andrei Radu}
\author{Emil Stan}
\affiliation{Institute of Space Sciences, P.O.Box MG 23, RO 77125, Bucharest-M\u{a}gurele, Romania}

\date{\today}

\begin{abstract}
We extended a previous qualitative study of the intermittent behaviour of a chaotical nucleonic system, 
by adding a few quantitative analyses: of the configuration and kinetic energy spaces, power spectra, 
Shannon entropies, and Lyapunov exponents. The system is regarded as a classical "nuclear billiard" 
with an oscillating surface of a 2D Woods-Saxon potential well. For the monopole and dipole vibrational 
modes we bring new arguments in favour of the idea that the degree of chaoticity increases when shifting 
the oscillation frequency from the adiabatic to the resonance stage of the interaction. The order-chaos%
-order-chaos sequence is also thoroughly investigated and we find that, for the monopole deformation case, 
an intermittency pattern is again found. Moreover, coupling between one-nucleon and collective degrees 
of freedom is proved to be essential in obtaining chaotic states.
\end{abstract}


\maketitle

\section{\label{intro}Introduction}
We begin by briefly reminding that a conjugated continuous effort has been made to relate the emergence 
of the collective energy dissipation through one and two-body nuclear processes with the chaotical 
behaviour of nuclear systems \cite{burgio-95,baldo-96,baldo-98,blocki-78,ring-80,speth-81,wong-82,%
grassberger-83,sieber-89,rapisarda-91,abul-magd-91,blocki-92,blumel-92,baldo-93,blocki-93,berry-93,%
ott-93,bauer-94,hilborn-94,blumel-94,drozdz-94,drozdz-95,bauer1-95,jarzynski-95,bulgac-95,atalmi-96a,%
atalmi-96b,papachristou-08,felea-01,felea-02,bordeianu-08a,bordeianu-08b,bordeianu-08c,felea-09a}.

A few options in choosing the collective oscillation frequencies have come into focus in the past 
years, in connection with the onset of chaoticity for "nuclear billiards". First of all, the issue 
of dissipation into thermal motion of the adiabatic collective vibrational energy of the potential 
well was treated for several multipolarities by Burgio, Baldo \textit{et al.} \cite{burgio-95,%
baldo-96,baldo-98}. On the other hand, when trying to associate different vibration frequencies to 
various nuclear processes, the path to chaos was found to be changed with the order of multipole 
\cite{felea-01,felea-02,bordeianu-08a,bordeianu-08b,bordeianu-08c,felea-09a}.

This paper was intended to bring a quantitative argumentation, based on a systematic study of the 
configuration and kinetic energy spaces, power spectra, informational entropies, and largest Lyapunov 
exponents. The study was done in completion of a few qualitative types of analysis previously presented 
\cite{felea-09a}: sensitive dependence on the initial conditions, single-particle phase space maps, 
fractal dimensions of Poincare maps, and autocorrelation functions.

In short, we remind that, by studying the nucleonic dynamics in a Woods-Saxon potential, one can 
find an increase of the chaotical degree of the system behaviour as raising the frequency of 2D 
wall oscillation.

The main result of \cite{felea-09a} was reported in relation with an intermittent route to chaos for 
the monopole vibrations close to the resonance phase of a nuclear interaction. Still, we mention 
that the purpose of these two coupled etudes was only to emphasize the detection of such intermission 
for the "nuclear billiards" and not to establish its type according to \cite{pomeau-80}, nor to compare 
it with other intermittency patterns from known experimental results \cite{berge-80,pomeau-81,linsay-81,%
testa-82,jeffries-82,dubois-83,yeh-83,huang-87,richetti-87,kreisberg-91}.

\section{\label{sec:1}Toy model}
We continue the study on a classical dynamical system proposed by Burgio, Baldo \textit{et al.} \cite{%
burgio-95,baldo-96,baldo-98}, system composed of a number of $A$ nucleons with no charge, spin, or 
internal structure. A two-dimensional deep Woods-Saxon potential well, regarded as "nuclear billiard", 
is periodically hit by the nucleons. The Bohr Hamiltonian in polar coordinates is a sum of two components: 
kinetic ($E_{kin.}$) and potential ($E_{pot.}$), the kinetic one decoupling into radial ($E_r$), centrifugal 
($E_L$), and collective terms ($E_{coll.}$):

\begin{equation}
E_{kin.} = E_r + E_L + E_{coll.} = {\sum_{j=1}^{A}}\left( \frac{p_{r_{j}}^{2}}{2m}+\frac{p_{\theta _{j}}^{2}}{2mr_{j}^{2}} \right)%
+\frac{p_{\alpha }^{2}}{2M},
\end{equation}

\begin{equation}
E_{pot.} = {\sum_{j=1}^{A}} V\left(r_{j},R\left( \theta _{j}\right) \right)+\frac{M\Omega ^{2}\alpha ^{2}}{2}.
\end{equation}

The phase space is defined by particle and collective momenta and their conjugate coordinates: 
$\left( r,p_{r}\right)$, $\left( \theta,p_{\theta}\right)$ and $\left( \alpha,p_{\alpha}\right)$. 
The collective coordinate $\alpha $ oscillates with $\Omega $ frequency, the Inglis mass $M$ is 
equal to $mAR_{0}^{2}$, and the nucleon mass: $m=938\ \rm{MeV}$.

For the time being we are not interested in studying the nucleon dynamics beyond the Woods-Saxon 
barrier:

\begin{equation}
V\left( r_j,R\left( \theta _j\right) \right) =\frac{V_0}{1+\exp \left[ \frac{r_j-R\left( \theta _j,\alpha \right) }a\right] },
\end{equation}

and therefore we choose a deep well: $V_{0}=-1500\ \rm{MeV}$ and accordingly, a low value for the 
diffusivity coefficient: $a=0.01\ \rm{fm}$.

When considering the two-dimensional case, the frontier of the collective motion is described as a 
function of the collective variable $\alpha $ and of the Legendre polynomials $P_{L}\left( \cos \theta_{j}\right)$ 
\cite{burgio-95,baldo-96,baldo-98,felea-09a}:

\begin{equation}
R_j = R\left( \theta _j,\alpha \right) =R_0\left[ 1+\alpha P_L\left( \cos \theta_j\right) \right].
\end{equation}

The oscillation degree of the potential well $L$ is considered for the monopole $\left(0\right)$, 
dipole $\left(1\right)$ and quadrupole case $\left(2\right)$.

If the surface has a stationary behaviour, or whenever one takes into account the uncoupled Hamilton 
equations (\rm{UCE}) for the particle:

\begin{equation}
\stackrel{\cdot }{r_j}=\frac{p_{r_j}}m,\ 
\stackrel{\cdot }{\theta _j}=\frac{p_{\theta _j}}{mr_j^2},\ 
\stackrel{\cdot }{p_{r_j}}=\frac{p_{\theta _j}^2}{mr_j^3}-\frac{\partial V}{\partial r_j},\ 
\stackrel{\cdot }{p_{\theta _j}}=-\frac{\partial V}{\partial R_j}\cdot \frac{\partial R_j}{\partial \theta _j},
\end{equation}

and collective degrees of freedom (\textit{d.o.f.}):

\begin{equation}
\stackrel{\cdot }{\alpha }=\frac{p_\alpha }M,\ 
\stackrel{\cdot }{p_\alpha}=-M\Omega ^2\alpha -\sum_{j=1}^{A}\left(\frac{\partial V}{\partial R_j} \cdot \frac{\partial R_j}{\partial \alpha }\right),
\end{equation}

$R_{0}$ has a fix value, chosen for consistency with previous papers \cite{burgio-95,baldo-96,baldo-98,felea-09a} 
as $6\ \rm{fm}$.

\begin{figure}
\resizebox{1.\hsize}{!}{\includegraphics{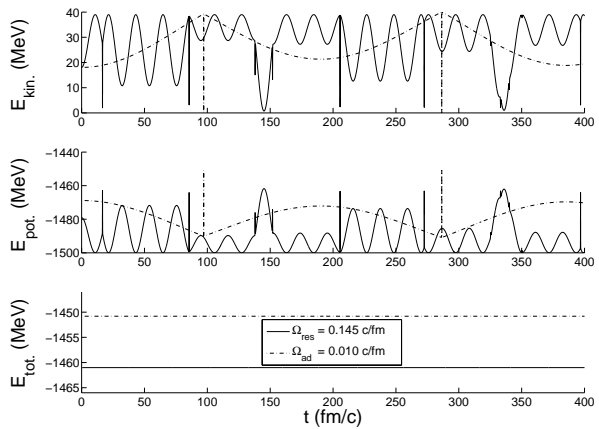} }
\caption{\label{fig:1}Energy conservation for the adiabatic and resonance phase of the interaction (L = 0).}
\end{figure}

A Runge-Kutta type algorithm (order 2-3) with an optimized step size was used for solving the system of 
differential equations, while keeping the absolute errors for the phase space variables under $10^{-6}$ 
and conserving the total energy with relative error: $\Delta E/E \approx 10^{-8}$ (Fig. \ref{fig:1}).

We imposed the equilibrium condition between the pressure exerted by the particles and the 
mechanical pressure of the wall \cite{burgio-95,baldo-96,baldo-98,felea-09a} and thus obtained for 
the initial equilibrium value of the collective variable, perturbed with a small value:

\begin{equation}
\alpha_{0}=\frac{-1+\sqrt{1+8T/{mR_0^2\Omega ^2}}}{2}+0.15,
\end{equation}

where $T=36\ \rm{MeV}$ \cite{burgio-95,baldo-96,baldo-98,felea-09a} is the two-dimensional kinetic 
energy and also the temperature of the nuclear system, when considering the natural system of units 
($\hbar=c=k_{B}=1$).

\section{\label{sec:2}The quantitative analysis of the route to chaos}
We carry on the study begun in \cite{felea-09a} by gradually changing the degree of vibration of the 
potential wall from a slow motion (adiabatic state) to a rapid one (the so-called resonance state 
of the interaction). In the first case, the collective motion is described by a radian frequency 
smaller than $0.05\ c/\rm{fm}$. The latter is dominated by frequencies close to the one-nucleon 
collisional frequency:

\begin{equation}
\omega _{part}=\frac{\pi}{R_0}\cdot\sqrt{\frac{2T}{m}}\approx 0.145\ c/\rm{fm}.
\end{equation}

The physical motivation for studying the one-nucleon chaotical dynamics in the "nuclear billiard", 
ranging the frequencies from the adiabatic to the resonance regime of a nuclear interaction, is 
explained in some detail in \cite{felea-09a}.

We briefly remind that the process of nuclear multifragmentation can be viewed as a resonance 
process and that for smaller excitation energies, nuclear evaporation or breakup of a projectile 
nucleus occurs when the energy is shared between the collective and one-nucleon degrees of freedom.

By using a few types of analyses: sensitive dependence on the initial conditions, single-particle 
phase space maps, fractal dimensions of Poincare maps and autocorrelation functions, we emphasized 
that an intermittent route to chaos is observed in the monopole case when increasing the 
vibrational frequency to $\Omega = 0.1\ c/\rm{fm}$ \cite{felea-09a}. In the resonance phase of the 
interaction the onset of chaotical behaviour was found to be earlier than at any other adiabatic 
oscillations of the Woods-Saxon potential well.

We present here other methods \cite{schuster-84} promoting the idea that the degree of chaoticity 
increases when moving from the adiabatic to the resonance regime: analyses of the configuration 
and kinetic energy spaces, power spectra, generalized informational entropies and Lyapunov exponents. 
Furthermore, we try to identify possible pathways to chaos, including the intermittent one, previously 
put in evidence \cite{felea-09a}.

\subsection{Configuration and kinetic energy spaces}

In order to establish if a specific physical system presents a chaotic dynamics and to identify 
possible routes to chaos we analyzed the behaviour of a small bunch of trajectories in the 
configuration and in the kinetic energy space, respectively. For example, we took five trajectories 
separated by an $\epsilon = \Delta r=0.01\ \rm{fm}$ aperture, while keeping the rest of the initial 
phase space variables constant and let the system evolve over a given time $\left(\Delta t = 1,600\ \rm{fm}/c\right)$.

\begin{figure*}[tbp]
\resizebox{1.\hsize}{!}{\includegraphics{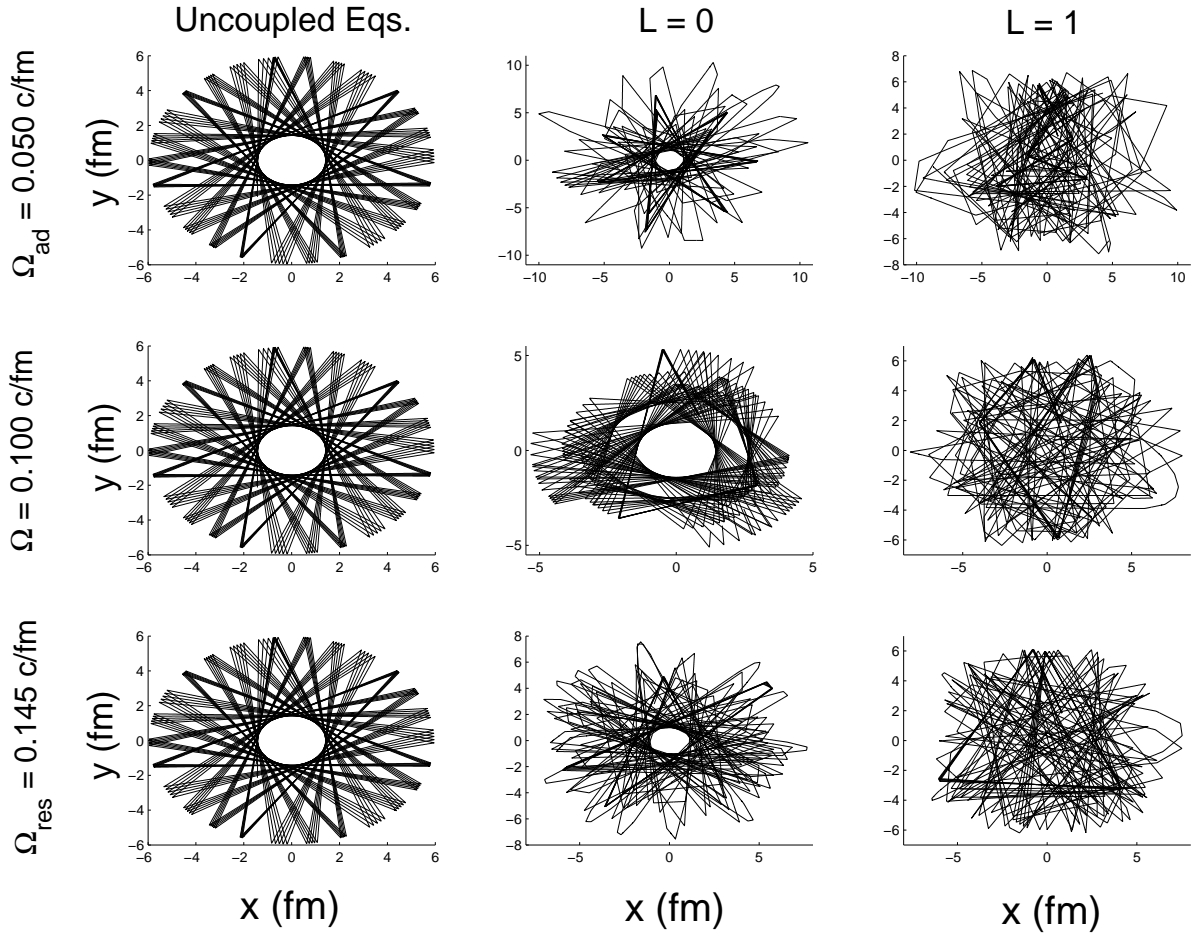} }
\caption{\label{fig:2}The variance in time $\left(\Delta t = 1,600\ \rm{fm}/c\right)$ of a packet of one-particle trajectories in the configuration space for adiabatic (upper row), intermediate (middle row), and resonance (lower row) collective vibrations for the uncoupled single and collective motion, monopole, and dipole cases.}
\end{figure*}

\begin{figure*}[tbp]
\resizebox{1.\hsize}{!}{\includegraphics{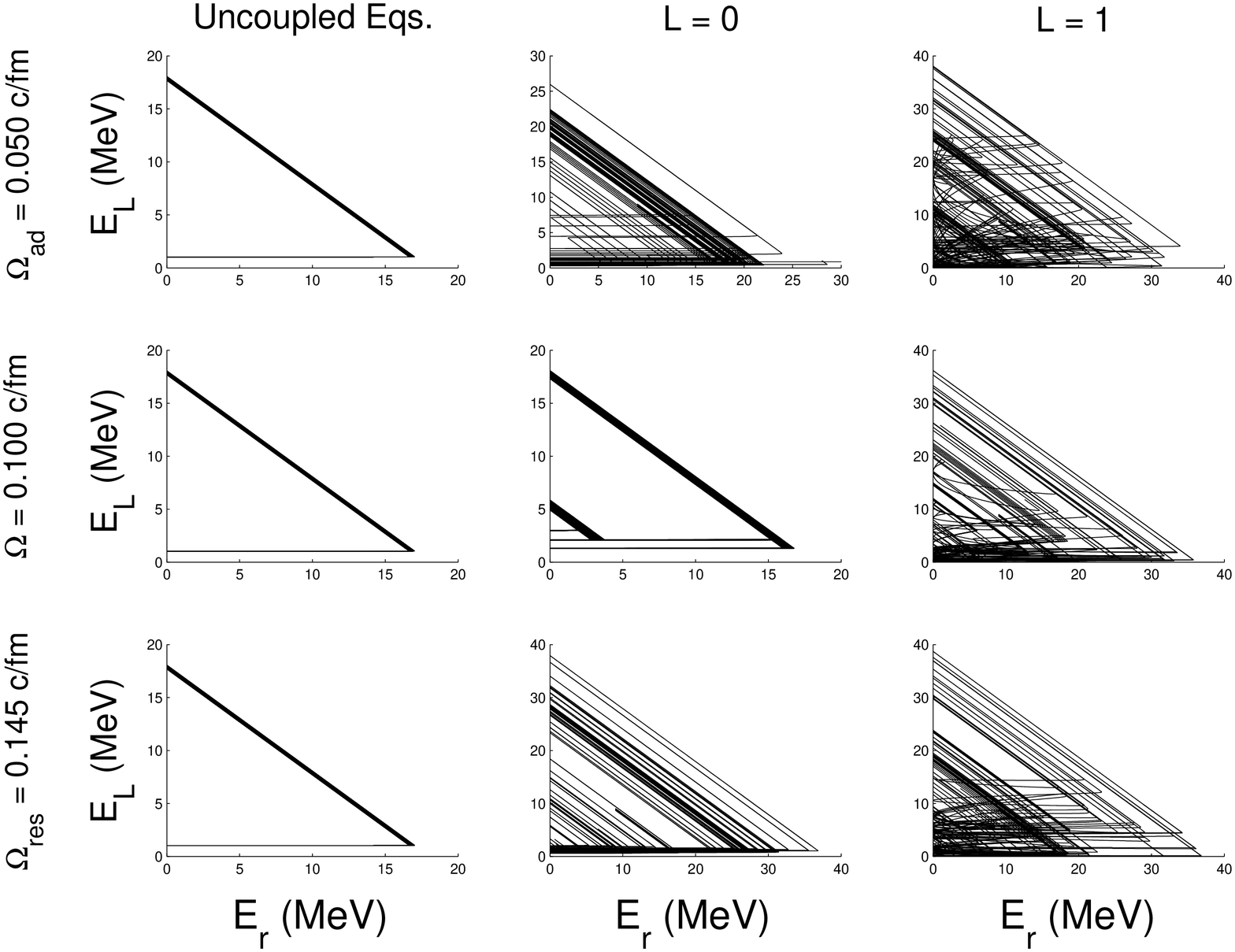} }
\caption{\label{fig:3}The variance in time $\left(\Delta t = 1,600\ \rm{fm}/c\right)$ of a packet of one-particle trajectories in the kinetic energy space for adiabatic (upper), intermediate (middle), and resonance (lower) collective vibrations for the uncoupled single and collective motion, monopole, and dipole cases.}
\end{figure*}

For the transient stages from adiabatic to resonance, the temporal evolution of an initially confined 
trajectory bundle was studied for the monopole and dipole oscillation modes of the potential wall 
and also for the limit situation, in which the individual and collective degrees of freedom remain 
uncoupled (Figs. \ref{fig:2} and \ref{fig:3}).

The configuration space revealed a high degree of symmetry in $\left(r,\theta\right)$ plane in 
both cases, $\left( 4+2\right)$ uncoupled nonlinear differential equations and monopole (left and 
middle panels). Also, that the central zone remained uncovered, reflecting the conservation of the 
nucleon angular momentum.

Another important conclusion was issued from the definition of the stability concept of a dynamical 
system. For the aforementioned cases the dispersed trajectory pack periodically regroups on the 
frontier that delimits the forbidden zone of the phase space. This type of behaviour corresponds 
with the definition of stability given by Poisson (for e.g., in \cite{holmes-96}). We will herewith 
remind that the Poisson stability defines as steady the movement of a particle system of which 
configuration comes close, from time to time, to the initial position.

At a first glance, on the simple \rm{UCE} case one can distinguish two extremities of the radius of 
the particle periodic motion in the 2D potential well: $r_{min}(UCE) \approx 1.42\ \rm{fm}$, and 
$r_{Max}(UCE) \approx 5.96\ \rm{fm}$. These values correspond to the roots of the boundary equation:

\begin{equation}
E = \frac{p_{\theta}^2}{2mr^2}+V\left(r \right),
\end{equation}

for a given one-nucleon energy $E$, when the radial component of velocity vanishes ($\stackrel{\cdot }{r}=0$).

The analysis of the particle motion in the configuration space is similar to that applied to any system 
with bound unclosed trajectories. The nucleonic motion takes place within a circular crown (the so-called 
\textit{annulus}) determined by the concentric circles of $r_{min}$ and $r_{Max}$ radii. The trajectory 
is symmetric about any turning point.

For $\Delta t = 1,600\ \rm{fm}/c$ we descry in the \rm{UCE} case as much as $27$ distinct apocenters 
(at $r=r_{Max}$). These are correlated with a number of $13$ complete and one incomplete revolutions 
about the center of the force field (\textit{i.e.} $27$ straight lines before $r(t)$ changes its sense 
of variation).

The particle completely sweeps over twice the 2D configuration space after $1,537\ \rm{fm}/c$. However, 
we notice that the bound trajectories are open, which means that the orbits never pass twice through 
a given point (see Fig. \ref{fig:2}), which is in concordance with Bertrand's theorem. We briefly 
remind that for a bound orbit to be closed, the angle between two consecutive apocenters must be:

\begin{equation}
\Delta \theta = 2\pi \cdot \frac{n_{1}}{n_{2}},
\end{equation}

\textit{i.e.} after $n_{2}$ revolutions about the center, the radius vector should sweep out a multiple 
$n_{1}$ of $2\pi$ \rm{radians}. For the \rm{UCE} case above considered we consequently obtain: 
$\Delta \theta_{UCE} \approx 4\pi/13$ \rm{radians} (Fig. \ref{fig:2} - left column).

The kinetic energy points are displayed in right isosceles triangular shaped patterns (Fig. \ref{fig:3}), 
whose hypotenuses are described by Eqs. (11) and (12), for the two specific non-chaotical situations: 
\rm{UCE} and the intermittent monopolar "window" emerged at $\Omega = 0.1\ c/\rm{fm}$:

\begin{equation}
E_{L_{UCE}} = 18.05 - E_r,
\end{equation}

\begin{equation}
E_{L_{\Omega=0.1\: c/\rm{fm}}} = 18.09 - E_r.
\end{equation}

Moreover, for the intermittency frequency of monopole oscillations, one can notice a second smaller 
segment with the same negative slope:

\begin{equation}
E_{L_{\Omega=0.1\: c/\rm{fm}}} = 5.87 - E_r.
\end{equation}

For the uncoupled differential equations there are a couple of extreme values for the centrifugal kinetic 
energy: $E_{L_{min}}(UCE) = 1.02\ \rm{MeV}$, and $E_{L_{Max}}(UCE) = 18.05\ \rm{MeV}$, associated with 
$r_{Max}(UCE)$ and respectively, with $r_{min}(UCE)$ (left column of Fig. \ref{fig:3} and Eq. (1)).

As for the monopolar intermittency, we can distinguish just five distinct values for the $E_{L}$: 
$1.32\ \rm{MeV}$, $2.13\ \rm{MeV}$, $3.01\ \rm{MeV}$, $5.87\ \rm{MeV}$, and $18.09\ \rm{MeV}$ (central 
plot of Fig. \ref{fig:3}), correlated with stationary radii: $r_{1} \approx 5.26\ \rm{fm}$, $r_{2} %
\approx 4.14\ \rm{fm}$, $r_{3} \approx 3.49\ \rm{fm}$, $r_{4} \approx 2.51\ \rm{fm}$, and $r_{5} \approx %
1.43\ \rm{fm}$ (Eq. (1), Fig. 2 of \cite{felea-09a}, and central plot of Fig. \ref{fig:2}). Thus, the 
nucleonic motion for the intermittent case is composed of alternated revolutions about the force field 
centre, forming a cyclic symmetrical structure, for e.g., $r_{1}$, $r_{5}$, $r_{2}$, $r_{4}$, $r_{3}$, %
$r_{4}$, $r_{2}$, $r_{5}$, $r_{1}$, and so on (Fig. \ref{fig:2}). This behaviour can be easily verified 
through the sensitivity dependence on the initial conditions analysis, previously presented (third column 
of Fig. 2 - \cite{felea-09a}). It should also be mentioned that, following this radius alternation, 
the nucleon covers in 2D configuration space $\approx 2\pi$ radians after $6$ full revolutions in almost 
$590\ \rm{fm}/c$.

Concluding, we highlight once more, that ordered, non-chaotical events, exhibit periodical symmetrical 
patterns in the configuration and kinetic energy spaces. This was shown to be a characteristic feature 
of the uncoupled nonlinear Hamilton equations case and also, of the steady, intermittent behaviour 
arisen in the monopole case at $0.1\ c/\rm{fm}$ vibrational radian frequency.

A tendency to compactly fill the kinetic energy space when increasing the monopolar vibrations (from 
$0.05\ c/\rm{fm}$ to $0.145\ c/\rm{fm}$) was observed, except for the intermittency situation above 
described. For the $L = 1$ oscillation mode of the potential well, it seems that at the same frequency 
($\Omega = 0.1\ c/\rm{fm}$) a somewhat intermittent behaviour could also come out, but this was proved 
to be elusive, as verified when reverting to this issue with the help of informational entropies and 
Lyapunov exponents and analyzing the system on longer time periods.

\subsection{Power spectra}

In order to better distinguish between a multiple periodical behaviour that can also exhibit an erratic 
pattern and chaos we used the Fourier transform of the analyzed signals:

\begin{equation}
x\left( \omega \right) = \lim_{T\rightarrow \infty} \int^{T}_{0} e^{i\omega t} \cdot x\left( t\right) \ dt\ ,
\end{equation}

\begin{equation}
x\left( \omega \right) = \lim_{N\rightarrow \infty} \sum^{N}_{n=0} e^{i\omega t_{n}} \cdot x\left( t_{n}\right).
\end{equation}

For a multiple periodical movement the power spectrum:

\begin{equation}
P\left( \omega \right) = \left|x\left( \omega \right)\right|^{2},
\end{equation}

will only contain a number of discrete lines: the fundamental frequencies of the system and their associated 
sets of harmonics, while the chaotical behaviour is completely aperiodical and is represented by a continuous 
or quasi-continuous broadband.

\begin{figure*}[tbp]
\resizebox{1.\hsize}{!}{\includegraphics{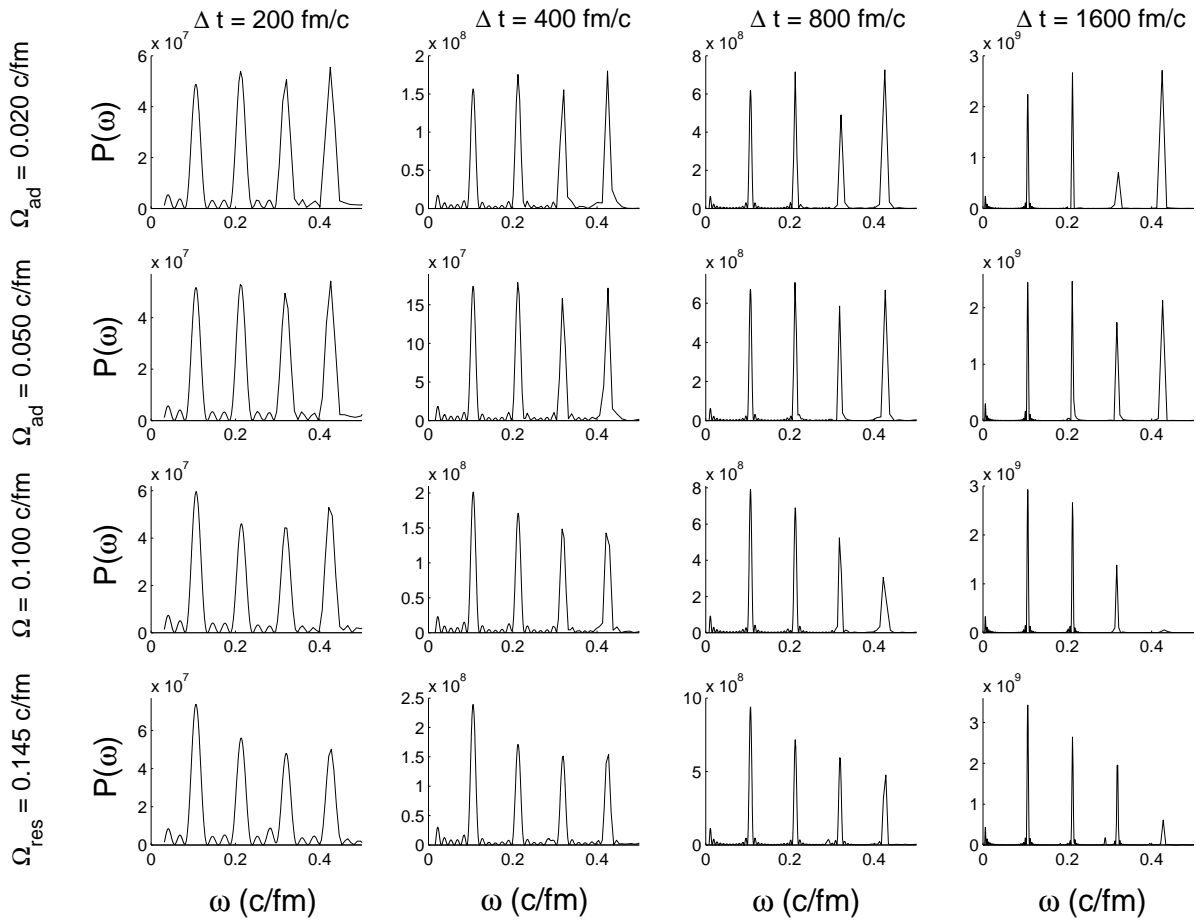} }
\caption{\label{fig:4}The power spectra of the one-nucleon radial variable for different phases of the interaction (from adiabatic to resonance regime) calculated for the uncoupled single and collective degrees of freedom case.}
\end{figure*}

\begin{figure*}[tbp]
\resizebox{1.\hsize}{!}{\includegraphics{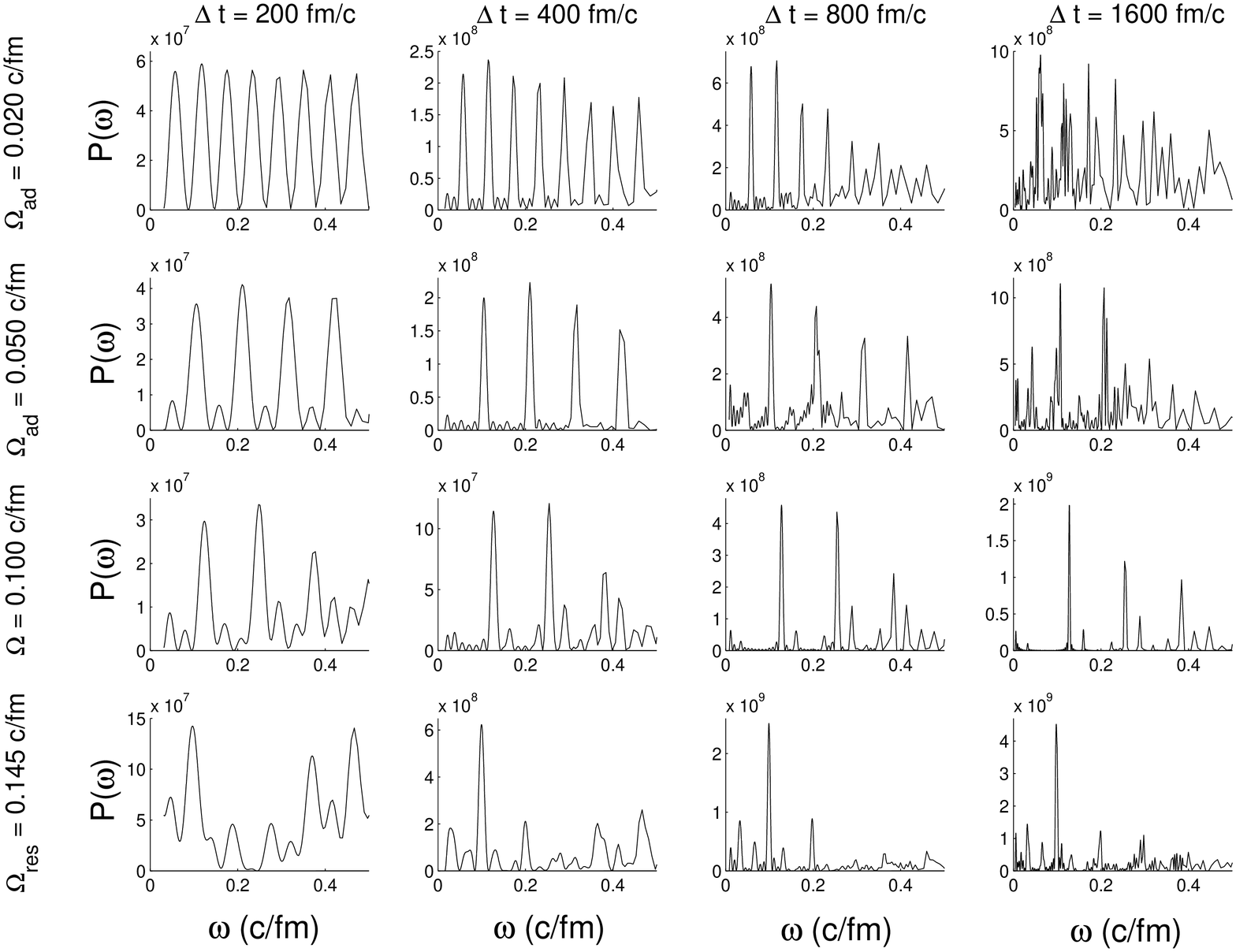} }
\caption{\label{fig:5}The power spectra of the one-nucleon radial variable for different phases of the interaction (from adiabatic to resonance regime) calculated for the monopole case.}
\end{figure*}

\begin{figure*}[tbp]
\resizebox{1.\hsize}{!}{\includegraphics{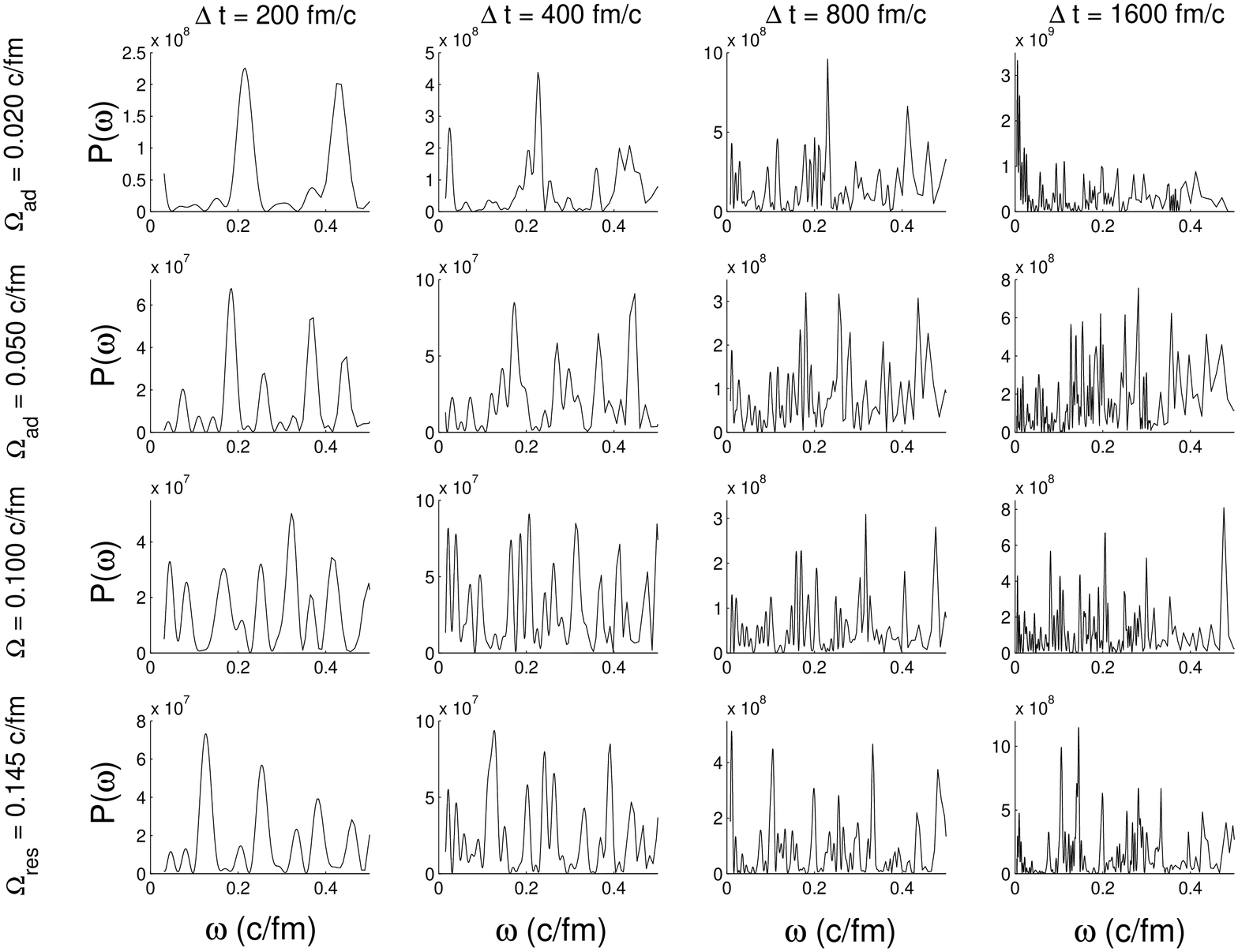} }
\caption{\label{fig:6}The power spectra of the one-nucleon radial variable for different phases of the interaction (from adiabatic to resonance regime) calculated for the dipole case.}
\end{figure*}

\begin{figure*}[tbp]
\resizebox{1.\hsize}{!}{\includegraphics{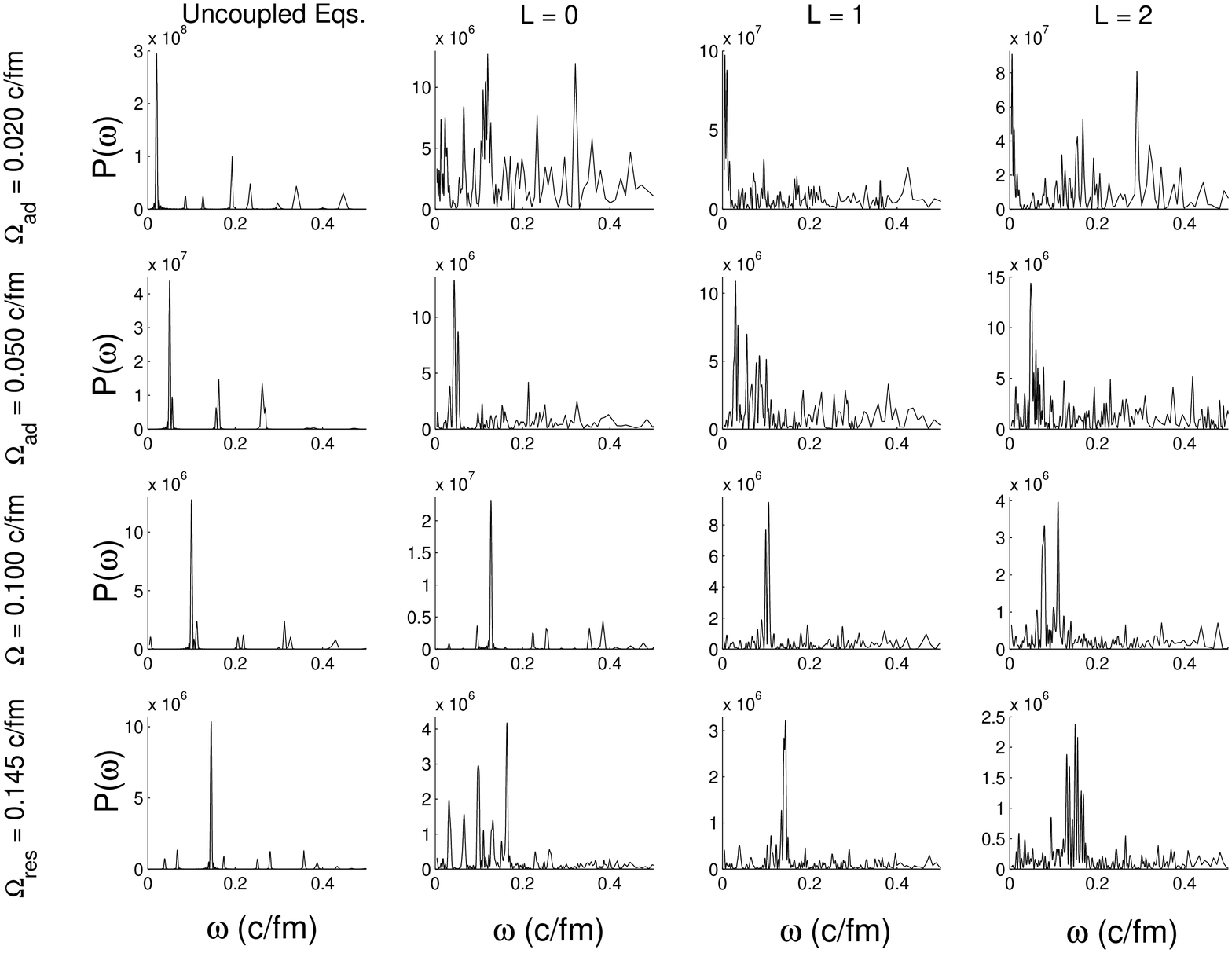} }
\caption{\label{fig:7}The power spectra of the collective coordinate for different phases of the interaction (from adiabatic to resonance regime) calculated for the uncoupled single and collective degrees of freedom, monopole, dipole and quadrupole cases.}
\end{figure*}

The obtained results are presented in Figures \ref{fig:4}-\ref{fig:7}. As a persistent feature of the 
physical system analyzed one should mention that for the monopole and dipole deformation degrees of 
the potential well (Figures \ref{fig:5} and \ref{fig:6}) the chaotic behaviour increases in time, thus 
confirming previous results.

The transition towards a chaotic regime was put again in evidence once passing from the adiabatic to the 
resonance stage of the interaction. The power spectra reveal, as expected, the intermittent feature of the 
transition in the monopolar case at $\Omega=0.1\ c/\rm{fm}$.

This can be detected for periods of time large enough ($\Delta t \geq 1,600\ \rm{fm}/c$) to positively 
identify chaotic patterns, by transition from a quasi-continuous spectrum of the one-nucleon radial 
coordinate ($\Omega _{ad}=0.02\ c/\rm{fm}$) to a discrete periodical one, containing fundamental 
frequencies of the system and its harmonics ($\Omega=0.1\ c/\rm{fm}$) and again to a continuous spectrum 
at the resonance vibrational frequency ($\Omega _{res}=0.145\ c/\rm{fm}$) (Fig. \ref{fig:5} - right panels). 
The order-chaos-order-chaos sequence can be also spotted out for the monopole oscillations in the power 
spectra of the collective degree of freedom (Fig. \ref{fig:7} - second column).

The temporal series of the radius variable show a symmetrical sawtooth waveform for the uncoupled 
situation at any chosen vibration frequency, and also for the monopole case at adiabatic collective 
oscillations. For the rest, in general an asymmetrical sawtooth form defines the series, but sometimes, 
more complicated patterns appear at higher multipole orders (Figures 1-4 - \cite{felea-09a}).

The difference between two successive maxima in the temporal series of the radius variable for the \rm{UCE} 
case is: $T_{0_{UCE}} \approx 1,537/26 = 59.1\ \rm{fm}/c$ (left column of Fig. \ref{fig:2}) and can be also 
obtained from the sensitive dependence on the initial conditions analysis (see Fig. 1 - \cite{felea-09a}). 
The fundamental frequency for the radial sawtooth temporal series: $\omega_{0_{UCE}}=2\pi/T_{0_{UCE}}=0.106\ c/\rm{fm}$ 
and its first three harmonics: $\omega_{1_{UCE}}=0.212\ c/\rm{fm}$, $\omega_{2_{UCE}}=0.318\ c/\rm{fm}$, 
and $\omega_{3_{UCE}}=0.424\ c/\rm{fm}$, can be easily traced down in Figure \ref{fig:4}.

As for the "window" of intermittency at $L = 0$, we obtained: $T_{0_{int}} \approx 590/12 = 49.2\ \rm{fm}/c$ 
(Fig. 2 - \cite{felea-09a} and central plot of current Fig. \ref{fig:2}). This gives the corresponding fundamental 
frequency: $\omega_{0_{int}}=0.128\ c/\rm{fm}$ and its associated harmonics: $\omega_{1_{int}}=0.256\ c/\rm{fm}$, 
$\omega_{2_{int}}=0.384\ c/\rm{fm}$, and $\omega_{3_{int}}=0.512\ c/\rm{fm}$ (Fig. \ref{fig:5}).

\subsection{Shannon entropies}

In order to further investigate route to chaos, we paid attention to the time evolution of the generalized 
informational entropy (or Shannon entropy), introduced as usually \cite{schuster-84,penrose-79,atalmi-98,%
kowalski-98,bialas-99,latora-99,latora-00}:

\begin{equation}
S_{Shannon}\left(t\right)=-{\sum^{N\left(t\right)}_{k=1}} p_k \cdot \ln p_k,
\end{equation}

$N\left(t\right)$ being the number of gradually occupied cells until the time $t$.

This type of entropy is actually a number which quantifies the time rate of information production for a 
chaotic trajectory \cite{ott-93}. We consider in the first place the case of a particle that at every 
moment occupies a cell of the two-dimensional lattice phase space with a $p_k$ probability:

\begin{equation}
p_k=1/N_{total\ cells},
\end{equation}

where:

\begin{equation}
N_{total\ cells}=N_{r} \cdot N_{p_{r}} \cdot N_{\theta},
\end{equation}

$N_{r}$, $N_{p_{r}}$, and $N_{\theta}$ are the number of bins of the $\left( r,p_{r},\theta\right)$ 
lattice. For $p_{\theta}$ is a constant of motion for the monopole and the \rm{UCE} cases, we use 
for comparisons only these three phase space variables.

As an alternative measure for the above defined entropy we also used the cumulative filling percentage of 
the one-nucleon phase space:

\begin{equation}
\eta\left(t\right)=\frac{N\left(t\right)}{N_{total\ cells}} \cdot 100\ \left(\%\right).
\end{equation}

\begin{figure}[tbp]
\resizebox{1.\hsize}{!}{\includegraphics{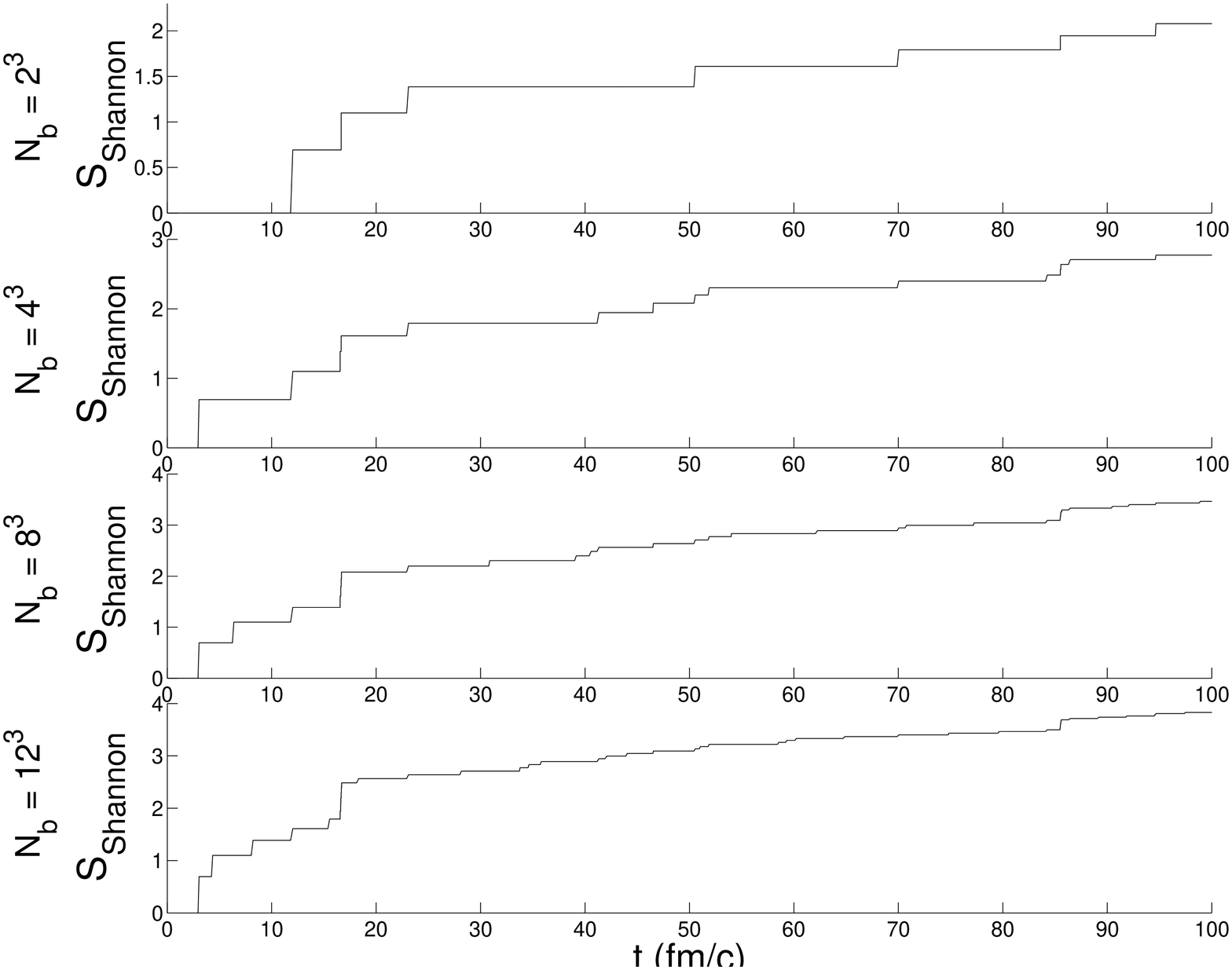} }
\caption{\label{fig:8}The Shannon entropy (upper panel) and the filling degree of the one-particle phase space (lower panel) for all frequencies studied ($\Omega = 0.02 - 0.145\ c/\rm{fm}$), computed for the uncoupled single and collective degrees of freedom case.}
\end{figure}

In the first place, for a given wall frequency of vibration and for a certain multipolarity (here, for 
$\Omega _{res}=0.145\ c/\rm{fm}$ and $L = 0$), we studied the dependence of the Shannon entropy with the 
number of bins. A clear tendency for smoothing the entropy curve was found when decreasing the bin. A 
reduced number of cells ($N_{b}=2^3$) is characterized by an entropy formed from a small number of 
high-amplitude Heaviside functions. As the number of bins increases (for e.g., here to $12^3$), 
the entropy gets a more realistic representation, being composed of a superior number of low-amplitude 
step functions (Fig. \ref{fig:8}).

Moreover, the filling percentage $\eta$ of the one-nucleon phase space maps can drastically differ with 
the size of the bin. Thus, after the system evolved over $400\ \rm{fm}/c$, a phase space with $8$ bins 
is entirely covered, $64$ bins can be filled in with $0.8594$ probability, a $26.95$ filling percentage 
for $512$ cells can be found, and we counted only as much as $226$ bins occupied out of a total of $1,728$ 
(\textit{i.e.} $\eta = 13.08\ \%$).

At a first glance one can identify a series of entropy plateaus, which could be put in correspondence 
with stationary or quasi-stationary thermodynamic values of the system if a large number of particles 
would be under study. Some of them will vanish when considering a large number of bins. However, those 
surviving for $N_{b} \rightarrow \infty$ could be associated with stationary nucleonic states in the 
chosen potential well in the limit of a large number of degrees of freedom.

\begin{figure}[tbp]
\resizebox{1.\hsize}{!}{\includegraphics{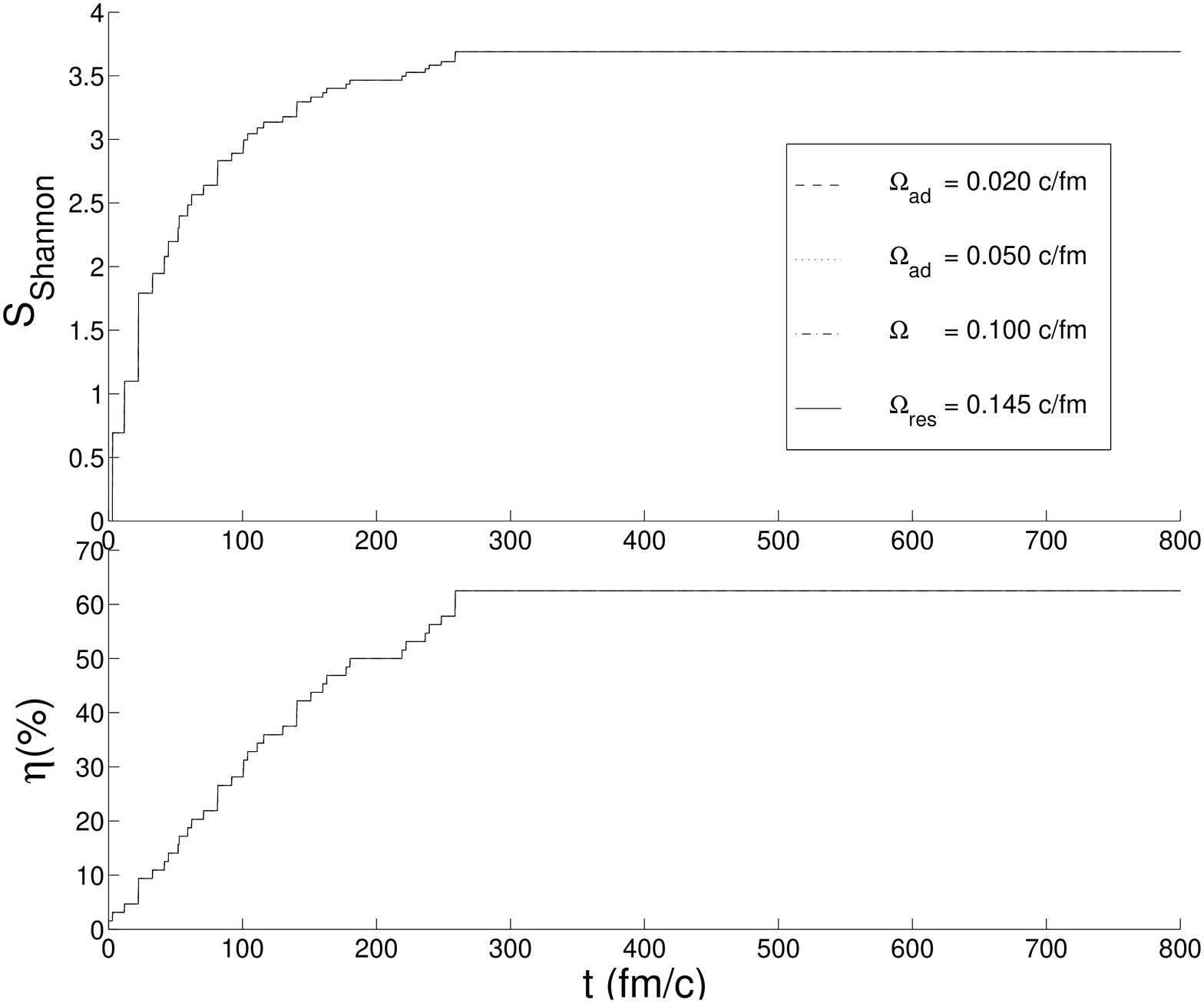} }
\caption{\label{fig:9}The Shannon entropy (upper panel) and the filling degree of the one-particle phase space (lower panel) for all frequencies studied ($\Omega = 0.02 - 0.145\ c/\rm{fm}$), computed for the uncoupled single and collective degrees of freedom case.}
\end{figure}

\begin{figure}[tbp]
\resizebox{1.\hsize}{!}{\includegraphics{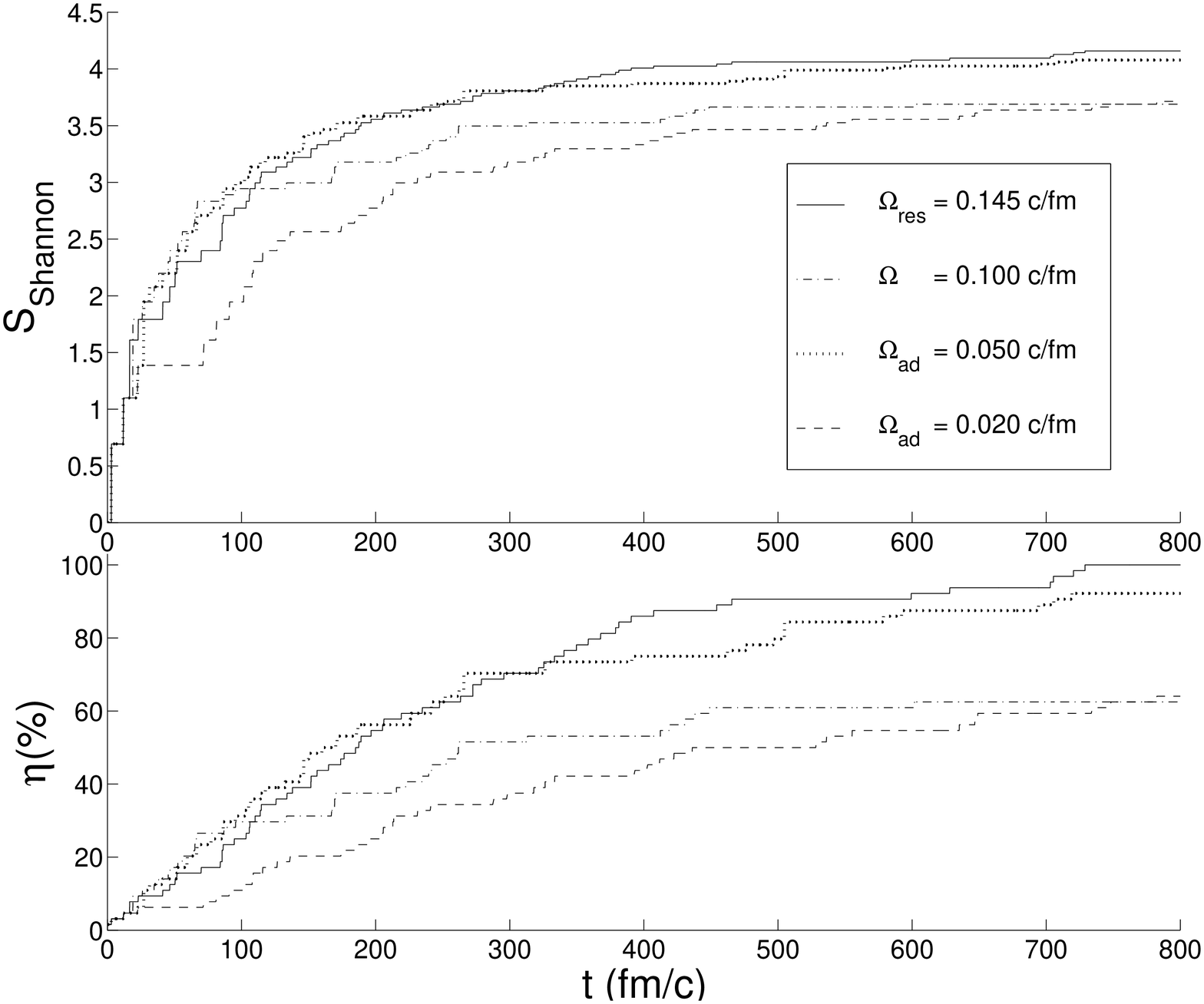} }
\caption{\label{fig:10}The Shannon entropy (upper panel) and the filling degree of the one-particle phase space (lower panel) for all frequencies studied ($\Omega = 0.02 - 0.145\ c/\rm{fm}$), computed for the first degree of multipolarity (L = 0).}
\end{figure}

\begin{figure}[tbp]
\resizebox{1.\hsize}{!}{\includegraphics{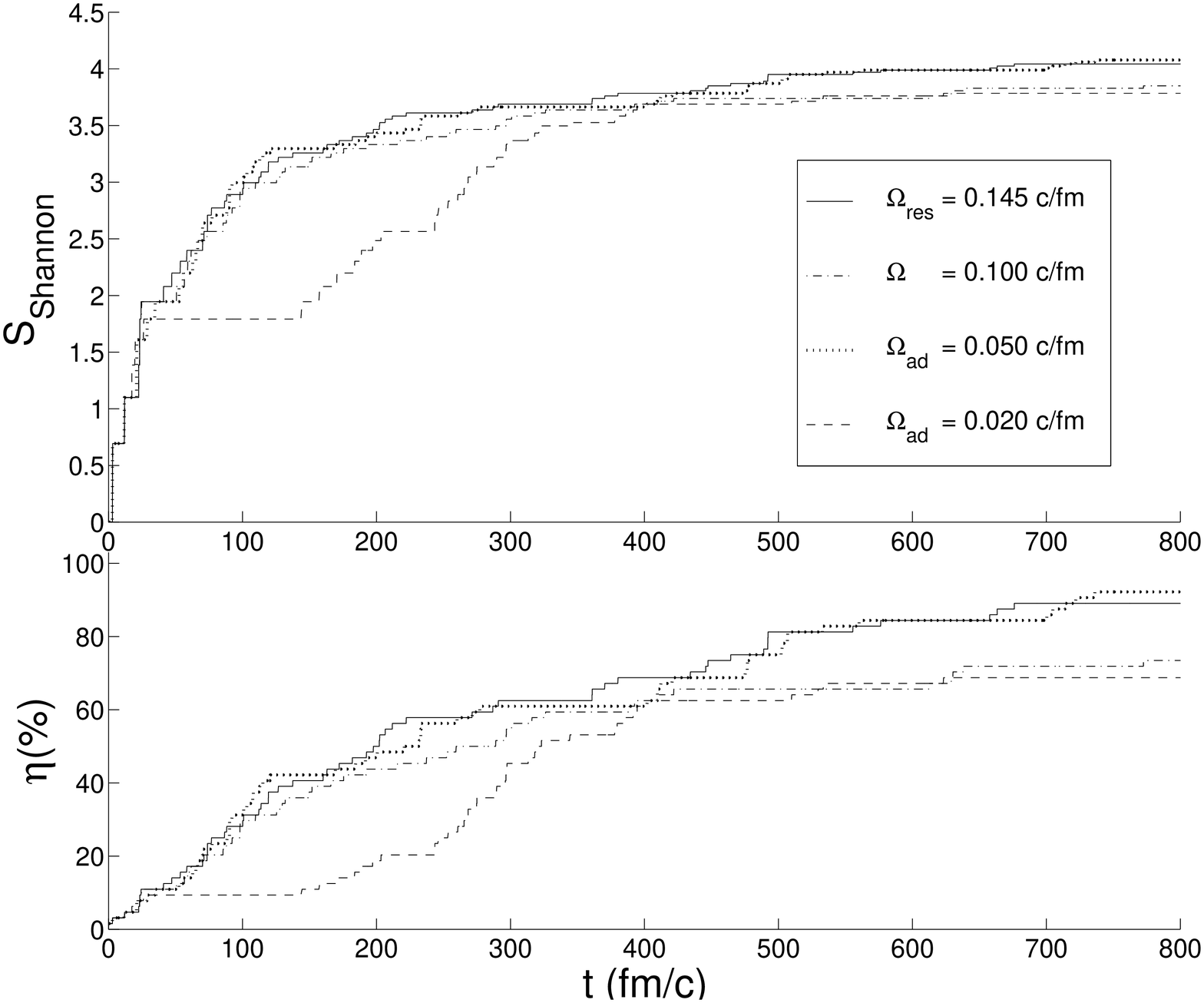} }
\caption{\label{fig:11}The Shannon entropy (upper panel) and the filling degree of the one-particle phase space (lower panel) for all frequencies studied ($\Omega = 0.02 - 0.145\ c/\rm{fm}$), computed for the dipole case (L = 1).}
\end{figure}

\begin{figure}[tbp]
\resizebox{1.\hsize}{!}{\includegraphics{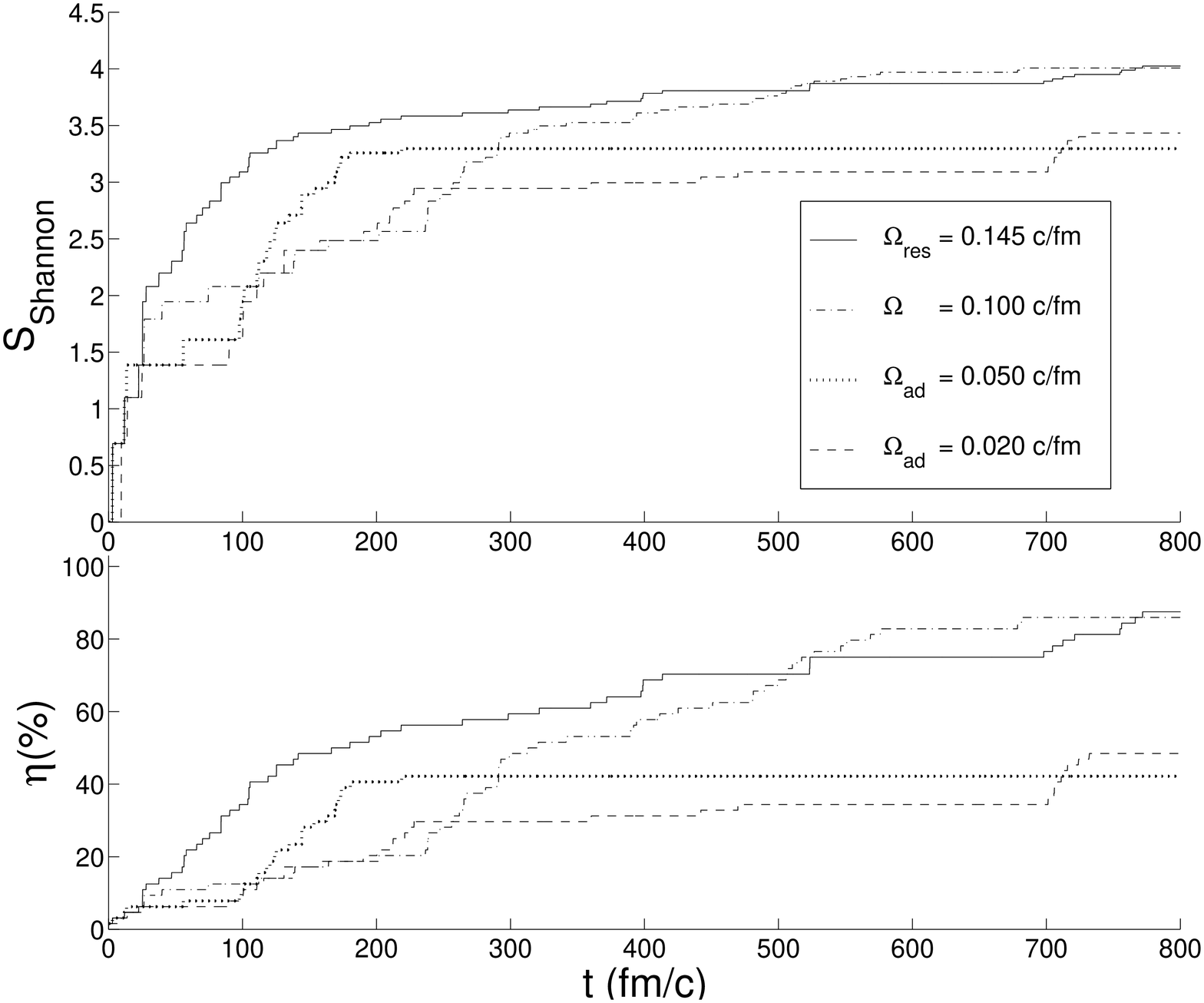} }
\caption{\label{fig:12}The Shannon entropy (upper panel) and the filling degree of the one-particle phase space (lower panel) for all frequencies studied ($\Omega = 0.02 - 0.145\ c/\rm{fm}$), computed for the quadrupole oscillations (L = 2).}
\end{figure}

For a given 2D phase space lattice formed of $N_{b}=4^3$ bins we present in Figures \ref{fig:9}-\ref{fig:12} 
a comparison between the informational entropies of the physical system in study, starting from the adiabatic 
stage of interaction and gradually increasing the vibrational wall frequency towards the dipole resonance 
value, $\Omega _{res}=0.145\ c/\rm{fm}$. The slopes for the resonance frequency case were found to be 
significantly higher than for the adiabatic one ($\Omega _{ad}=0.02\ c/\rm{fm}$) for all multipolarities 
involved.

Another comparison revealed significant differences between the onset times of the quasi-constant Shannon 
entropy values for all cases taken into consideration. Thus, for four vibrational radian frequencies 
and for four coupling modes of the Hamilton equations we show the informational entropy values after 
$800\: \rm{fm}/c$ (Table \ref{tab:1}) and the associated phase space filling degrees (Table \ref{tab:2}). 
Also, in Table \ref{tab:3}, are presented the periods of time after which the filling percentages $\eta$ 
equal unity.

\begin{table}
\caption{\label{tab:1}The computed $S_{Shannon}\left(t=800\: \rm{fm}/c\right)$ of the $\left( r\leftrightarrow p_{r}\leftrightarrow\theta\right)$ one-particle phase space maps at several multipolarities and frequencies of wall vibration}
\begin{ruledtabular}
\begin{tabular}{ccccc}
\hline\noalign{\smallskip}
Oscillation frequency & \rm{UCE} & $L=0$ & $L=1$ & $L=2$ \\
\noalign{\smallskip}\hline\noalign{\smallskip}
$\Omega \:_{ad} = 0.020\: c/\rm{fm}$ & $3.6889$ & $3.7136$ & $3.7842$ & $3.4340$ \\
$\Omega \:_{ad} = 0.050\: c/\rm{fm}$ & $3.6889$ & $4.0775$ & $4.0775$ & $3.2958$ \\
$\Omega \:\;\ \   = 0.100\: c/\rm{fm}$ & $3.6889$ & $3.6889$ & $3.8501$ & $4.0073$ \\
$\Omega _{res} = 0.145\: c/\rm{fm}$ & $3.6889$ & $4.1589$ & $4.0431$ & $4.0254$ \\
\noalign{\smallskip}\hline
\end{tabular}
\end{ruledtabular}
\end{table}

\begin{table}
\caption{\label{tab:2}The filling percentage $\eta$ of the $\left( r\leftrightarrow p_{r}\leftrightarrow\theta\right)$ one-particle phase space maps at several multipolarities and frequencies of wall vibration}
\begin{ruledtabular}
\begin{tabular}{ccccc}
\hline\noalign{\smallskip}
Oscillation frequency & \rm{UCE} & $L=0$ & $L=1$ & $L=2$ \\
\noalign{\smallskip}\hline\noalign{\smallskip}
$\Omega \:_{ad} = 0.020\: c/\rm{fm}$ & $62.50$ & $64.06$ & $68.75$ & $48.44$ \\
$\Omega \:_{ad} = 0.050\: c/\rm{fm}$ & $62.50$ & $92.19$ & $92.19$ & $42.19$ \\
$\Omega \:\;\ \   = 0.100\: c/\rm{fm}$ & $62.50$ & $62.50$ & $73.44$ & $85.94$ \\
$\Omega _{res} = 0.145\: c/\rm{fm}$ & $62.50$ & $100.00$ & $89.06$ & $87.50$ \\
\noalign{\smallskip}\hline
\end{tabular}
\end{ruledtabular}
\end{table}

\begin{table}
\caption{\label{tab:3}The time (in $\rm{fm}/c$) at which the informational entropies of the $\left( r\leftrightarrow p_{r}\leftrightarrow\theta\right)$ one-particle phase space maps at several multipolarities and frequencies of wall vibration have the maximum value (\textit{i.e.} $\eta  = 100\ \%$)}
\begin{ruledtabular}
\begin{tabular}{ccccc}
\hline\noalign{\smallskip}
Oscillation frequency & \rm{UCE} & $L=0$ & $L=1$ & $L=2$ \\
\noalign{\smallskip}\hline\noalign{\smallskip}
$\Omega \:_{ad} = 0.020\: c/\rm{fm}$ & $>10^{5}$ & $6,023$ & $6,359$ & $5,356$ \\
$\Omega \:_{ad} = 0.050\: c/\rm{fm}$ & $>10^{5}$ & $1,618$ & $4,223$ & $>10^{5}$ \\
$\Omega \:\;\ \   = 0.100\: c/\rm{fm}$ & $>10^{5}$ & $11,442$ & $3,241$ & $2,758$ \\
$\Omega _{res} = 0.145\: c/\rm{fm}$ & $>10^{5}$ & $729$ & $1,887$ & $10,571$ \\
\noalign{\smallskip}\hline
\end{tabular}
\end{ruledtabular}
\end{table}

\begin{figure}[tbp]
\resizebox{1.\hsize}{!}{\includegraphics{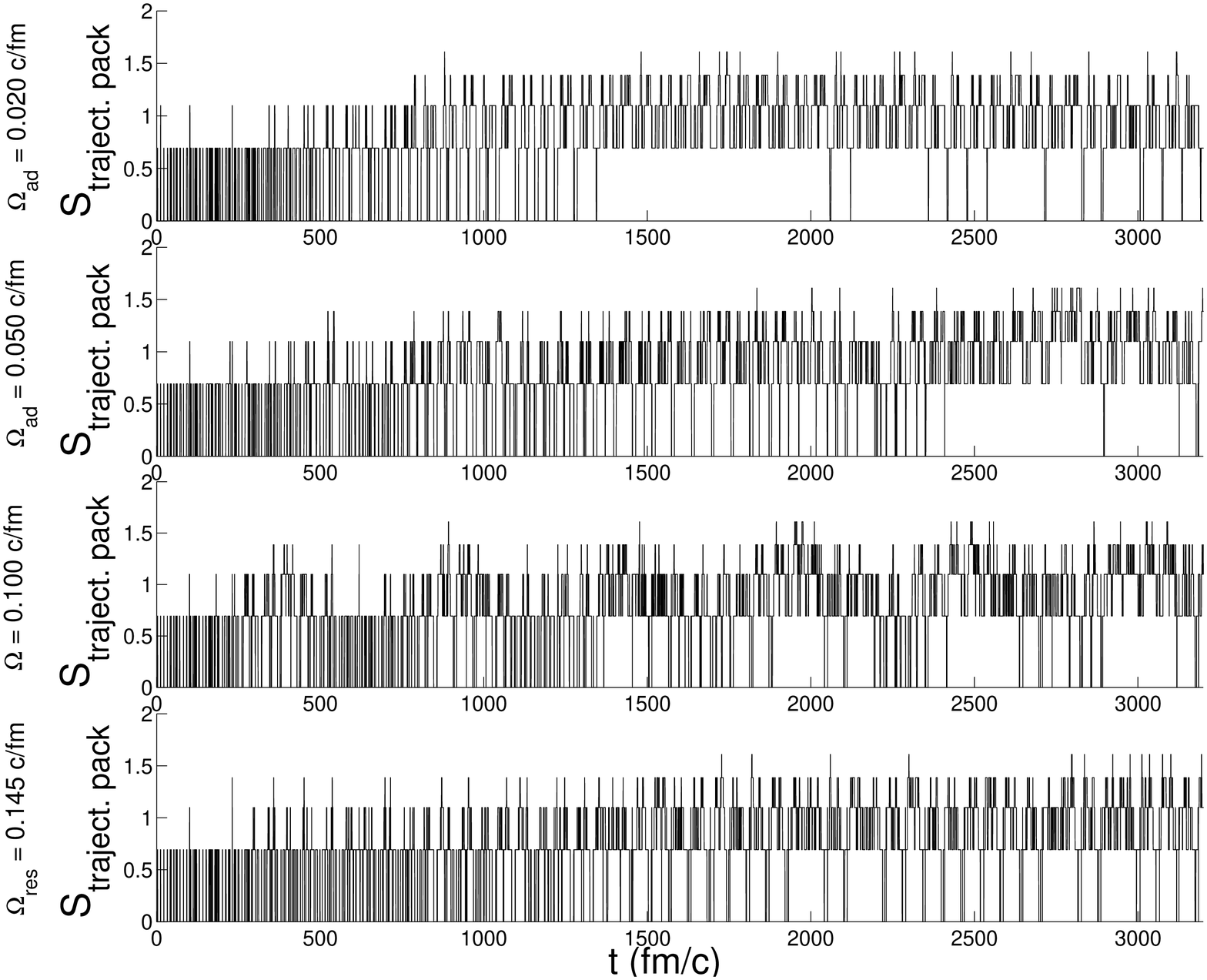} }
\caption{\label{fig:13}The Shannon entropy of a bunch of five one-particle close trajectories for frequencies between $0.02$ and $0.145\ c/\rm{fm}$ (uncoupled single and collective degrees of freedom case).}
\end{figure}

\begin{figure}[tbp]
\resizebox{1.\hsize}{!}{\includegraphics{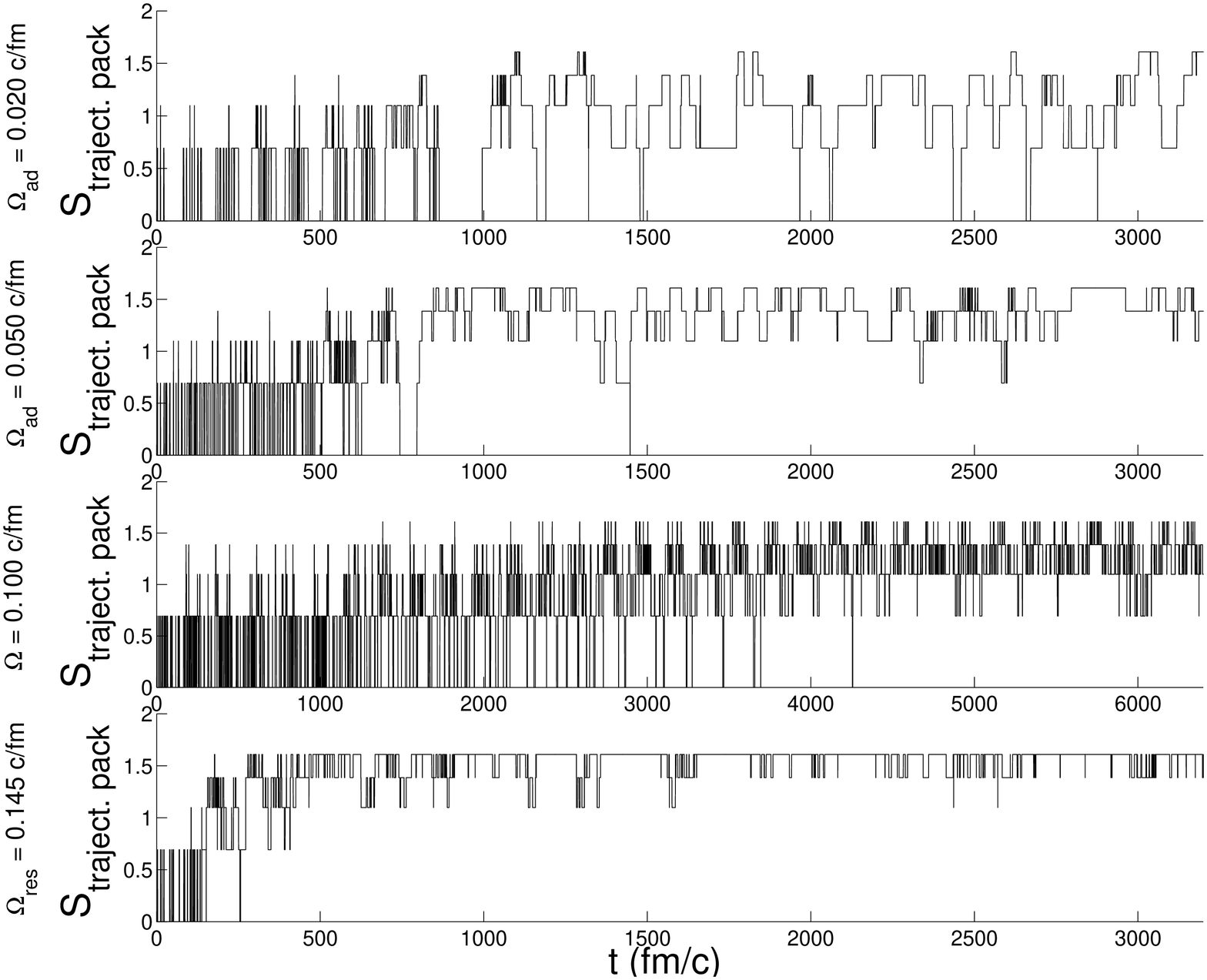} }
\caption{\label{fig:14}The Shannon entropy of a bunch of five one-particle close trajectories for frequencies between $0.02$ and $0.145\ c/\rm{fm}$ (L = 0).}
\end{figure}

\begin{figure}[tbp]
\resizebox{1.\hsize}{!}{\includegraphics{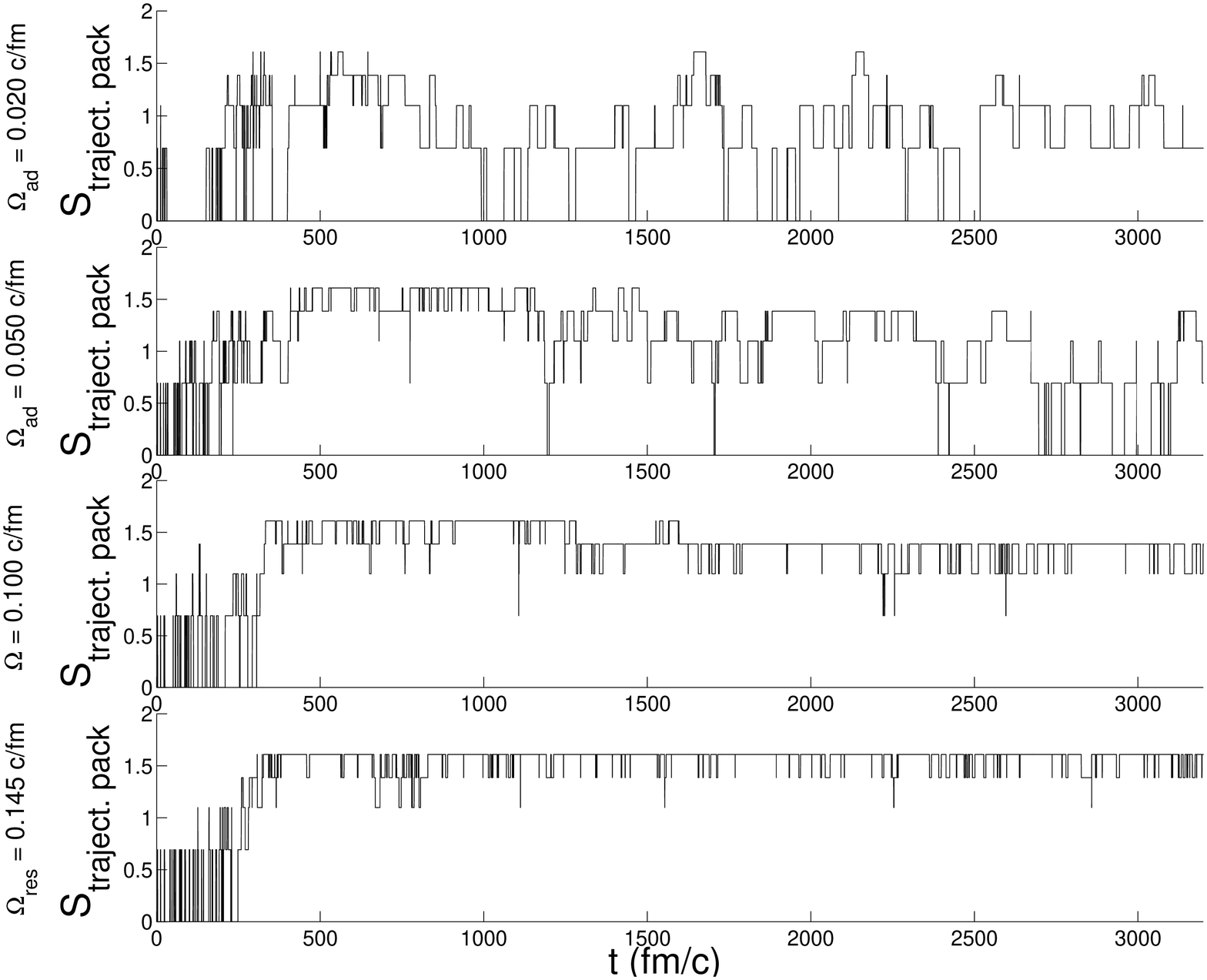} }
\caption{\label{fig:15}The Shannon entropy of a bunch of five one-particle close trajectories for frequencies between $0.02$ and $0.145\ c/\rm{fm}$ (L = 1).}
\end{figure}

\begin{figure}[tbp]
\resizebox{1.\hsize}{!}{\includegraphics{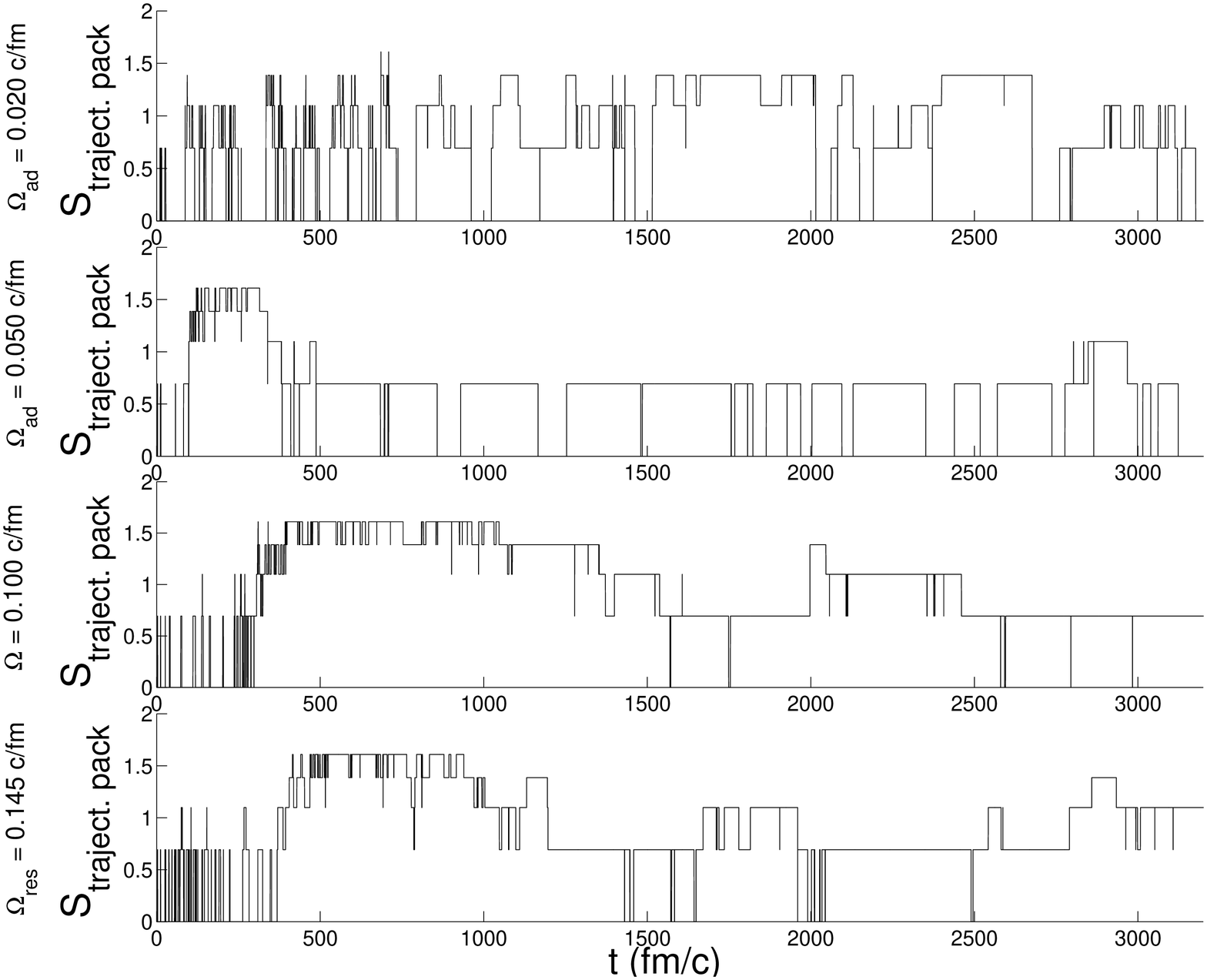} }
\caption{\label{fig:16}The Shannon entropy of a bunch of five one-particle close trajectories for frequencies between $0.02$ and $0.145\ c/\rm{fm}$ (L = 2).}
\end{figure}

We continue the analysis by further defining the Shannon entropy for a group of $w$ nearby orbits:

\begin{equation}
S_{traject.\ pack}\left(t\right)=ln\ N_{w}\left(t\right),
\end{equation}

so that the number of occupied cells is:

\begin{equation}
1 \leq N_w(t) \leq w,
\end{equation}

thus describing the spread of the trajectories at each moment of time $t$ (Figs. \ref{fig:13}-\ref{fig:16}). 
When reaching the maximum divergence, the entropy for five distinct phase space paths gets its highest 
value (Table \ref{tab:4}).

\begin{table}
\caption{\label{tab:4}The time (in $\rm{fm}/c$) at which the one-particle Shannon entropies of a pack of $w=5$ close orbits begin having the maximum value (\textit{i.e.} $S_{traject.\ pack} = 1.60944$) for various coupling degrees between the one-nucleon and the collective \textit{d.o.f.} and for the standard chosen wall frequencies}
\begin{ruledtabular}
\begin{tabular}{ccccc}
\hline\noalign{\smallskip}
Oscillation frequency & \rm{UCE} & $L=0$ & $L=1$ & $L=2$ \\
\noalign{\smallskip}\hline\noalign{\smallskip}
$\Omega \:_{ad} = 0.020\: c/\rm{fm}$ & $>10^{4}$ & $1,095$ & $555$ & $688$ \\
$\Omega \:_{ad} = 0.050\: c/\rm{fm}$ & $>10^{4}$ & $855$ & $476$ & $122$ \\
$\Omega \:\;\ \   = 0.100\: c/\rm{fm}$ & $>10^{4}$ & $4,133$ & $333$ & $395$ \\
$\Omega _{res} = 0.145\: c/\rm{fm}$ & $>10^{4}$ & $279$ & $327$ & $469$ \\
\noalign{\smallskip}\hline
\end{tabular}
\end{ruledtabular}
\end{table}

We begin the analysis with the \rm{UCE} case. The single and collective uncoupled \textit{d.o.f.} give 
birth to a quasi-laminar behaviour with a weak development of chaotic states. The one-particle 
informational entropy shows an identical evolution, no matter the frequency chosen. The orbit covers, 
after $800\ \rm{fm}/c$, only $62.50\ \%$ of the entire lattice (Table \ref{tab:2} and Figure \ref{fig:9}) 
and does not reach $100\ \%$, even after $\Delta t = 100,000\ \rm{fm}/c$ (Table \ref{tab:3}). Also, 
the phase space is not covered up by all five trajectories for the whole range of $10,000\ \rm{fm}/c$ 
considered, when analyzing $S_{traject.\ pack}$ (Fig. \ref{fig:13} and Table \ref{tab:4}).

For the dipole oscillations mode, at $\Omega_{ad} = 0.05\ c/\rm{fm}$, it appears that, after only 
$800\ \rm{fm}/c$, the entropy closes in upon its maximum value: $S_{Max} = ln\ N_{total\ cells} = 4.1589$ 
(Fig. \ref{fig:11} and Table \ref{tab:1}). However, on long periods of time, the real tendency is towards 
filling up the nucleonic phase space as rapid as the vibrational frequency is increased (Table \ref{tab:3}). 
The exact pattern is repeated when studying the Shannon entropy for closeby nucleonic trajectories (Fig. 
\ref{fig:15} and Table \ref{tab:4}).

We found quite the same feature for the monopole case, with exception for the intermittent "window" 
at $\Omega=0.1\ c/\rm{fm}$ (Tables \ref{tab:1}, \ref{tab:2} and Fig. \ref{fig:10}). The occupying rate 
is so small in the intermittent zone, that just at $11,442\ \rm{fm}/c$, the particle would have covered 
the whole phase space (see Table \ref{tab:3}). A similar conclusion can be drawn from Table \ref{tab:4} 
and Fig. \ref{fig:14} (with a double temporal scale scanned for the intermittent frequency). The trajectory 
pack informational entropy reaches its highest value after the longest one-particle evolution time of 
all: $4,133\ \rm{fm}/c$.

The quadrupole oscillation also reveals an apparent intermittent pattern, this time at $\Omega_{ad} = %
0.05\ c/\rm{fm}$. We call it intermittent because after $800\ \rm{fm}/c$ the nucleon fills in only 
$42.19\ \%$ of the total number of bins (Figure \ref{fig:12} and Table \ref{tab:2}), and a longer time 
than $100,000\ \rm{fm}/c$ is required to get to $\eta  = 100\ \%$ (Table \ref{tab:3}). However, this 
behaviour can be a misleading one, the Shannon entropy for a trajectory bunch showing exactly the opposite 
(see Fig. \ref{fig:16} and Table \ref{tab:4}), after $122\ \rm{fm}/c$ the orbits being completely dispersed.

\subsection{Lyapunov exponents}

We furthermore presented another quantitative analysis: the temporal evolution of the Lyapunov exponents, 
$\lambda\left(t\right)$. As previously shown, initial adjacent points in the phase space $\Delta x_{0} (t=0)$, 
can generate in time separated trajectories $\Delta x\left(t\right)$. When studying the evolution of a 
single phase space parameter, the one-dimensional Lyapunov exponent takes the form:

\begin{equation}
\lambda\left(t\right)=\lim_{\left|\Delta x_{0}\right|\rightarrow 0} ln\left|\frac{\Delta x\left(t\right)}{\Delta x_{0}}\right|.
\end{equation}

The generalization for obtaining the multi-dimensional Lyapunov exponent is then straightforward:

\begin{equation}
\lambda\left(t\right)=ln\frac{\left(\sum_{k=1}^{m}\left[x_{k}\left(t\right)-x_{k0}\right]^{2}\right)^{\frac{1}{2}}}{0.01},
\end{equation}

where the sum is taken over all $m = 4$ squared differences between final $x_{k}\left(t\right)$ and 
initial $x_{k0}$ one-nucleon phase space variables. Integration times of the order of $10^{3}$ \rm{fm}/c 
exclude errors when computing the Lyapunov exponents.

In short, we here remind that the trajectories can be classified as function of the Lyapunov exponents. 
Thus, one can distinguish periodical behaviours, for $\lambda=0$, dissipative movements with a fixed point 
or a basin of attraction $\left(\lambda<0\right)$, and aperiodical chaotic states $\left(\lambda>0\right)$, 
when the iterative discrete evolution of the solution series (Eqs. 5 and 6) leads to a chaotic pattern.

Another way of measuring the system sensitivity to initial conditions is to compute the largest Lyapunov 
exponent (\rm{LLE}). Usually a couple of methods can be employed, one based on the time dependence of the 
multi-dimensional Lyapunov exponent, the other on Wolf's standard method that uses a Gram-Schmidt 
Reorthonormalization of the tangent vectors \cite{wolf-85}. In the latter, the \rm{LLE} is obtained by 
taking the asymptotic value of the multi-dimensional Lyapunov exponent:

\begin{equation}
\lambda_{Max}=\lim_{t \rightarrow \infty} \frac{\lambda\left(t\right)}{t}.
\end{equation}

Still, this method has the disadvantage that the integration times have to be at least an order of 
magnitude larger than those here considered. Other methods are slightly less efficacious, being more 
CPU time-consuming when simulating strong chaotic systems \cite{ramasubramanian-00}.

\begin{figure*}[tbp]
\resizebox{1.\hsize}{!}{\includegraphics{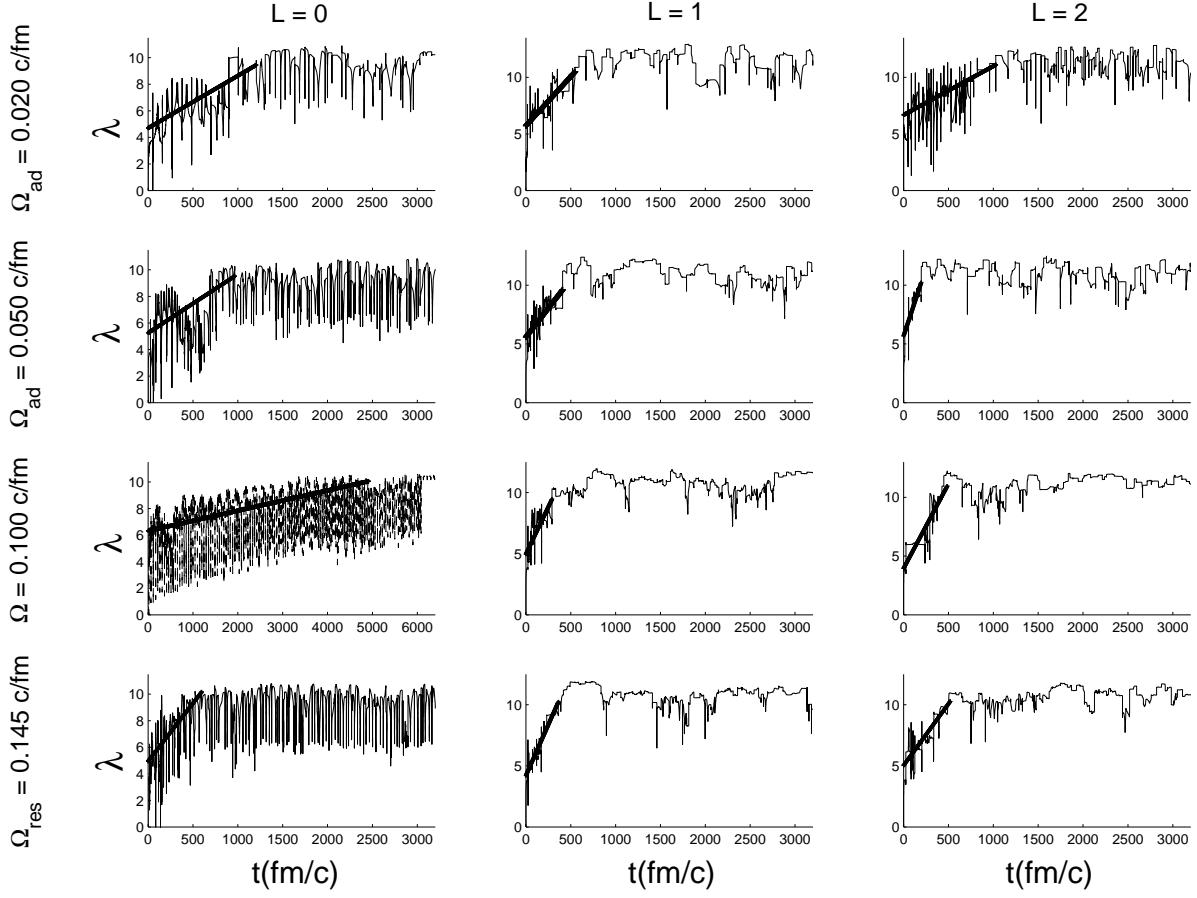} }
\caption{\label{fig:17}The temporal evolution of the Lyapunov exponents of the single-particle \textit{d.o.f.} for multiple radian frequencies $\Omega$ and multipolarities $L$ of the nuclear surface. The \rm{LLE} are given by fits (thickened solid lines) of the increasing portions of the 4-dimensional Lyapunov exponents.}
\end{figure*}

We consequently used the first method and noticed the saturation behaviour, \textit{i.e.} the arising of 
a plateau after a certain time $t_{c}$ (Fig. \ref{fig:17}). The straight lines represent fits whose slopes 
match the \rm{LLE} (Table \ref{tab:5}). They are in inverse proportion with the onset times of chaoticity 
$(\tau = 1/\lambda_{Max})$, being a measure of the trajectory decoupling at a microscopic level.

\begin{table}
\caption{\label{tab:5}The largest Lyapunov exponents (in $c/\rm{fm}$) computed as slopes of the increasing branches of the ordinary Lyapunov exponents for the degrees of multipole and vibrational frequencies used}
\begin{ruledtabular}
\begin{tabular}{cccc}
\hline\noalign{\smallskip}
Oscillation frequency & $L=0$ & $L=1$ & $L=2$ \\
\noalign{\smallskip}\hline\noalign{\smallskip}
$\Omega \:_{ad} = 0.020\: c/\rm{fm}$ & $0.003939$ & $0.008689$ & $0.004306$ \\
$\Omega \:_{ad} = 0.050\: c/\rm{fm}$ & $0.004432$ & $0.009829$ & $0.023203$ \\
$\Omega \:\;\ \   = 0.100\: c/\rm{fm}$ & $0.000761$ & $0.015454$ & $0.014402$ \\
$\Omega _{res} = 0.145\: c/\rm{fm}$ & $0.008739$ & $0.016662$ & $0.010086$ \\
\noalign{\smallskip}\hline
\end{tabular}
\end{ruledtabular}
\end{table}

When the single-particle and collective \textit{d.o.f.} remain uncoupled, the Lyapunov exponents 
basically oscillate between two quasi-stationary regimes. This happens for all vibrational frequencies 
involved, reflecting a periodical regrouping of orbits in two basins of attraction. The phase space 
not being covered, even after a hundred of thousand of $\rm{fm}/c$, computing the \rm{LLE} becomes 
futile for this case.

One can remark for dipole oscillations (Fig. \ref{fig:17} - middle column) a faster evolution towards 
reaching saturation states of the 4-dimensional Lypaunov exponents, once passing from the adiabatic 
$(\tau _{ad}=115\ \rm{fm}/c)$ to the resonance phase of the interaction $(\tau _{res}=60\ \rm{fm}/c)$.

In the monopolar case the intermittency can be easily traced at $0.1\ c/\rm{fm}$ vibrational frequency 
(Fig. \ref{fig:17} - left panels). During the intermittent stage, independent nearby orbits microscopically 
diverge with the slowest rate of all: $\tau =1,314\ \rm{fm}/c$ (Table \ref{tab:5}). In order to catch 
the heaving in sight of the stationary plateau at $\approx 4,917\ \rm{fm}/c$, the temporal scale was 
scanned over $6,400\ \rm{fm}/c$.

The study of the quadrupole collective oscillation case confirms the results obtained with all previous 
analyses. Namely, the neighbouring trajectories deviate one from each other after just $43\ \rm{fm}/c$ 
at an adiabatic frequency: $0.05\ c/\rm{fm}$. Also, when increasing $\Omega$, the \rm{LLE} evolution 
pattern exactly matches that found with informational entropy measured for a group of orbits (Tables 
\ref{tab:4} and \ref{tab:5}).

\section{\label{sec:3}Conclusions}
We investigated the chaotic nucleonic behaviour in a two-dimensional deep Woods-Saxon potential well 
for specific phases of the nuclear interaction. By comparing the order-to-chaos transition for these 
cases of interest, from adiabatic to resonance regime, it was shown that the couplings between the 
one-particle dynamics and high multipole vibrational modes significantly decrease the onset of the 
chaotic nucleonic motion towards realistic nuclear interaction time scale.

The quantitative study enfolded a plethora of analyses, pointing out that the paths to chaos for the 
"nuclear billiard" are dissimilar for the studied multipolarities. For the first two multipole degrees 
we noticed a more rapid emergence of chaotic states as moving on towards higher radian frequencies of 
oscillation. When analyzing the system with quadrupole collective deformations of the potential well, 
an order-strong chaos-weak chaos-order sequence is revealed. Still, as emphasized in the "Shannon 
entropies" subsection, the quadrupole case represents an intricate one, and further analysis would be 
required before concluding it.

Every type of quantitative analysis strengthened previous results regarding the monopolar intermittency 
route to chaos for the "nuclear billiard". The collective oscillation frequency for the intermittent 
behaviour was located prior to the resonance state of interaction (at $\Omega =0.1\ c/\rm{fm}$).

Further studies along the above issues are currently in progress. The used formalism can be improved by 
adding spin and charge to the nucleons. A semi-quantal treatment of this problem, including Pauli 
blocking effect, is hoped to shed more light on the discussed issue in the near future.

\begin{acknowledgments}
We wish to thank to R.I. Nanciu, I.S. Zgur\u{a}, A.\c{S}. C\^{a}rstea, G. P\u{a}v\u{a}la\c{s}, S. Zaharia, 
A. Ghea\c{t}\u {a}, M. Rujoiu, A. Mitru\c{t}, and R. M\u{a}rginean for fruitful discussions on this paper.
\end{acknowledgments}


\begin{thebibliography}{}
\bibitem{burgio-95} G.F. Burgio, M. Baldo, A. Rapisarda, and P. Schuck, Phys. Rev. C \textbf{52}, 2475 (1995).

\bibitem{baldo-96} M. Baldo, G.F. Burgio, A. Rapisarda, and P. Schuck, in \textit{Proceedings of the $XXXIV$ International Winter Meeting on Nuclear Physics, Bormio, Italy, 1996}, edited by I. Iori. arXiv:nucl-th/9602030

\bibitem{baldo-98} M. Baldo, G.F. Burgio, A. Rapisarda, and P. Schuck, Phys. Rev. C \textbf{58}, 2821 (1998).

\bibitem{blocki-78} J. Blocki, Y. Boneh, J.R. Nix, J. Randrup, M. Robel, A.J. Sierk, and W.J. Swiatecki, Ann. Phys. (N.Y.) \textbf{113}, 330 (1978).

\bibitem{ring-80} P. Ring and P. Schuck, \textit{The Nuclear Many Body Problem} (Springer-Verlag, Berlin, 1980) p. 388.

\bibitem{speth-81} J. Speth and A. van der Woude, Rep. Prog. Phys. \textbf{44}, 719 (1981).

\bibitem{wong-82} C.Y. Wong, Phys. Rev. C \textbf{25}, 1460 (1982).

\bibitem{grassberger-83} P. Grassberger and I. Procaccia, Phys. Rev. Lett. \textbf{50}, 346 (1983).

\bibitem{sieber-89} M. Sieber and F. Steiner, Physica D \textbf{44}, 248 (1990).

\bibitem{rapisarda-91} A. Rapisarda and M. Baldo, Phys. Rev. Lett. \textbf{66}, 2581 (1991).

\bibitem{abul-magd-91} A.Y. Abul-Magd and H.A. Weidenm\"{u}ller, Phys. Lett. B \textbf{261}, 207 (1991).

\bibitem{blocki-92} J. Blocki, F. Brut, T. Srokowski, and W.J. Swiatecki, Nucl. Phys. A\textbf{545}, 511c (1992).

\bibitem{blumel-92} R. Bl\"{u}mel and J. Mehl, J. Stat. Phys. \textbf{68}, 311 (1992).

\bibitem{baldo-93} M. Baldo, E.G. Lanza, and A. Rapisarda, Chaos \textbf{3}, 691 (1993).

\bibitem{blocki-93} J. Blocki, J.J. Shi, and W.J. Swiatecki, Nucl. Phys. A\textbf{554}, 387 (1993).

\bibitem{berry-93} M.V. Berry and J.M. Robbins, Proc. R. Soc., London, Sect. A \textbf{442}, 641 (1993).

\bibitem{ott-93} E. Ott, \textit{Chaos in Dynamical Systems} (Cambridge University Press, Cambridge, England, 1993).

\bibitem{bauer-94} W. Bauer, D. McGrew, V. Zelevinsky, and P. Schuck, Phys. Rev. Lett. \textbf{72}, 3771 (1994).

\bibitem{hilborn-94} R. Hilborn, \textit{Chaos and Nonlinear Dynamics} (Oxford University Press, Oxford, England, 1994).

\bibitem{blumel-94} R. Bl\"{u}mel and B. Esser, Phys. Rev. Lett. \textbf{72}, 3658 (1994).

\bibitem{drozdz-94} S. Drozdz, S. Nishizaki, and J. Wambach, Phys. Rev. Lett. \textbf{72}, 2839 (1994).

\bibitem{drozdz-95} S. Drozdz, S. Nishizaki, J. Wambach, and J. Speth, Phys. Rev. Lett. \textbf{74}, 1075 (1995).

\bibitem{bauer1-95} W. Bauer, D. McGrew, V. Zelevinsky, and P. Schuck, Nucl. Phys. A\textbf{583}, 93 (1995).

\bibitem{jarzynski-95} C. Jarzynski, Phys. Rev. Lett. \textbf{74}, 2937 (1995).

\bibitem{bulgac-95} A. Bulgac and D. Kusnezov, Chaos, Solitons and Fractals \textbf{5}, 1051 (1995).

\bibitem{atalmi-96a} A. Atalmi, M. Baldo, G.F. Burgio, and A. Rapisarda, Phys. Rev. C \textbf{53}, 2556 (1996). arXiv:nucl-th/9509020

\bibitem{atalmi-96b} A. Atalmi, M. Baldo, G.F. Burgio, and A. Rapisarda, in \textit{Proceedings of the $XXXIV$ International Winter Meeting on Nuclear Physics, Bormio, Italy, 1996}, edited by I. Iori. arXiv:nucl-th/9602039

\bibitem{papachristou-08} P.K. Papachristou, E. Mavrommatis, V. Constantoudis, F.K. Diakonos, and J. Wambach, Phys. Rev. C \textbf{77}, 044305 (2008). arXiv:nucl-th/0803.3336

\bibitem{felea-01} D. Felea, C. Be\c{s}liu, R.I. Nanciu, Al. Jipa, I.S. Zgur\u{a}, R. M\u{a}rginean, M. Haiduc, A. Ghea\c{t}\u{a}, and M. Ghea\c{t}\u{a}, in \textit{Proceedings of the $7^{th}$ International Conference "Nucleus-Nucleus Collisions", Strasbourg, 2000}, edited by W. Norenberg \textit{et al.} (North-Holland, Amsterdam, The Netherlands, 2001) p. 222.

\bibitem{felea-02} D. Felea, \textit{The Study of Nuclear Fragmentation Process in Nucleus-Nucleus Collisions at Energies higher than 1 A GeV}, Ph.D. thesis, University of Bucharest, Faculty of Physics (2002) p. 134.

\bibitem{bordeianu-08a} C.C. Bordeianu, C. Be\c{s}liu, Al. Jipa, D. Felea, and I.V. Grossu, Comput. Phys. Commun. \textbf{178}, 788 (2008).

\bibitem{bordeianu-08b} C.C. Bordeianu, D. Felea, C. Be\c{s}liu, Al. Jipa, and I.V. Grossu, Comput. Phys. Commun. \textbf{179}, 199 (2008).

\bibitem{bordeianu-08c} C.C. Bordeianu, D. Felea, C. Be\c{s}liu, Al. Jipa, and I.V. Grossu, Rom. Rep. in Phys. \textbf{60}, 287 (2008).

\bibitem{felea-09a} D. Felea, I.V. Grossu, C.C. Bordeianu, C. Be\c{s}liu, Al. Jipa, A.A. Radu, C.M. Mitu, and E. Stan, "Intermittency route to chaos for the nuclear billiard - a qualitative study", Phys. Rev. C (submitted).

\bibitem{pomeau-80} Y. Pomeau and P. Manneville, Commun. Math. Phys. \textbf{74}, 189 (1980).

\bibitem{berge-80} P. Berge, M. Dubois, P. Manneville, and Y. Pomeau, J. Phys. (Paris) \textbf{41}, L344 (1980).

\bibitem{pomeau-81} Y. Pomeau, J.C. Roux, A. Rossi, S. Bachelart, and C. Vidal, J. Phys. (Paris) \textbf{42}, L271 (1981).

\bibitem{linsay-81} P.S. Linsay, Phys. Rev. Lett. \textbf{47}, 1349 (1981).

\bibitem{testa-82} J. Testa, J. Perez, and C. Jeffries, Phys. Rev. Lett. \textbf{48}, 714 (1982).

\bibitem{jeffries-82} C. Jeffries and J. Perez, Phys. Rev. A \textbf{26}, 2117 (1982).

\bibitem{dubois-83} M. Dubois, M.A. Rubio, and P. Berge, Phys. Rev. Lett. \textbf{51}, 1446 (1983).

\bibitem{yeh-83} W.J. Yeh and Y.H. Kao, Appl. Phys. Lett. \textbf{42}, 299 (1983).

\bibitem{huang-87} J.Y. Huang and J.J. Kim, Phys. Rev. A \textbf{36}, 1495 (1987).

\bibitem{richetti-87} P. Richetti, P. DeKepper, J.C. Roux, and H.L. Swinney, J. Stat. Phys. \textbf{48}, 977 (1987).

\bibitem{kreisberg-91} N. Kreisberg, W.D. McCormick, and H.L. Swinney, Physica D \textbf{50}, 463 (1991).

\bibitem{schuster-84} H.G. Schuster, \textit{Deterministic Chaos: an introduction} (Physik-Verlag, Weinheim, Federal Republic of Germany, 1984).

\bibitem{holmes-96} P. Holmes and F. Diacu, \textit{Intalniri ceresti - originea haosului si a stabilitatii} (Societatea Stiinta si Tehnica SA, Bucuresti, Romania, 1996) p. 150.

\bibitem{penrose-79} O. Penrose, Rep. Prog. Phys. \textbf{42}, 129 (1979).

\bibitem{atalmi-98} A. Atalmi, M. Baldo, G. F. Burgio, and A. Rapisarda, Phys. Rev. C \textbf{58}, 2238 (1998).

\bibitem{kowalski-98} A.M. Kowalski, M.T. Martin, J. Nu\~{n}ez, A. Plastino, and A.N. Proto, Phys. Rev. A \textbf{58}, 2596 (1998).

\bibitem{bialas-99} A. Bialas, in \textit{Proceedings of the NATO-ASI International Summer School "Particle Production Spanning MeV and TeV Energies", Nijmegen, 1999}, edited by W. Kittel, P.J. Mulders, and O. Scholten, NATO Science Series C: Mathematical and Physical Sciences - Vol. 554 (Kluwer Academic Publishers, Nijmengen, The Netherlands, 1999).

\bibitem{latora-99} V. Latora and M. Baranger, Phys. Rev. Lett. \textbf{82} 520 (1999).

\bibitem{latora-00} V. Latora, M. Baranger, A. Rapisarda, and C. Tsallis, Phys. Lett. A \textbf{273}, 97 (2000).

\bibitem{wolf-85} A. Wolf, J.B. Swift, H.L. Swinney, and J.A. Vastano, Physica D \textbf{16}, 285 (1985).

\bibitem{ramasubramanian-00} K. Ramasubramanian and M.S. Sriram, Physica D \textbf{139}, 72 (2000).
\end{thebibliography}
\end{document}